\documentclass[12pt,preprint]{aastex}
\usepackage{epsf}
\usepackage{lscape}

\usepackage{graphicx}
\usepackage{natbib}
\def\mas{mag arcsec$^{-2}$}

\begin{document}

\title {THE FAINT END OF THE LUMINOSITY FUNCTION AND LOW SURFACE BRIGHTNESS GALAXIES}
\author {Margaret J. Geller} 
\affil{Smithsonian Astrophysical Observatory,
\\ 60 Garden St., Cambridge, MA 02138}
\email{mgeller@cfa.harvard.edu}
\author{Antonaldo Diaferio}
\affil{Dipartimento di Fisica Generale 'Amedeo Avogadro', 
\\Universit\`a degli Studi di Torino, via P. Giuria 1, 10125 Torino, Italy}
\affil{ INFN, Sezione di Torino, via P. Giuria 1, 
\\10125 Torino, Italy}
\affil{ Harvard-Smithsonian Center for Astrophysics,
\\60 Garden Street, Cambridge, MA 02138}
\email{adiaferio@cfa.harvard.edu}
\author {Michael J. Kurtz} 
\affil{Smithsonian Astrophysical Observatory,
\\ 60 Garden St., Cambridge, MA 02138}
\email{mkurtz@cfa.harvard.edu}
\author {Ian P. Dell'Antonio} 
\affil{Department of Physics, Brown University, 
\\Box 1843, Providence, RI 02912}
\email{ian@het.brown.edu}
\author {Daniel G. Fabricant} 
\affil{Smithsonian Astrophysical Observatory,
\\ 60 Garden St., Cambridge, MA 02138}
\email{dfabricant@cfa.harvard.edu}
\vfill \eject
\begin {abstract}
SHELS (Smithsonian Hectospec Lensing Survey) is a dense redshift survey covering a 4 square degree region to a limiting R = 20.6. In the  construction of the galaxy catalog and in the acquisition of spectroscopic targets, we paid careful attention to the survey completeness for lower surface brightness dwarf galaxies. Thus, although the survey covers a small area, it is a robust basis for computation of the slope of the faint end of the galaxy luminosity function to a limiting M$_{\rm R} = -13.3 + 5logh$.
We calculate the faint end slope in the R-band for  the subset of SHELS galaxies with redshifts in the range 0.02 $\leq z < 0.1$, SHELS$_{0.1}$. This sample contains 532 galaxies
with R$< 20.6$ and with a median surface brightness within the half light radius of SB$_{50,R}$ = 21.82 mag arcsec$^{-2}$. We used this sample to make one of the few direct measurements of  the dependence of the faint end of the galaxy luminosity function on surface brightness. For the sample as a whole the faint end slope, $\alpha = -1.31 \pm 0.04$, is consistent with
both the Blanton et al. (2005b) analysis of the SDSS and the Liu et al. (2008) analysis
of the COSMOS field. This consistency is impressive given the very different approaches of these three surveys. A magnitude limited sample of 135 galaxies with optical spectroscopic redshifts with mean half-light surface brightness, SB$_{50,R} \geq 22.5$ mag arcsec$^{-2}$
is unique to SHELS$_{0.1}$. The faint end slope is $\alpha_{22.5} = -1.52\pm 0.16$. 
SHELS$_{0.1}$ shows that lower surface brightness objects dominate the faint end slope of the luminosity function in the field, underscoring the importance of surface brightness limits in evaluating measurements of the faint end slope and its evolution.
\end{abstract}
\keywords {galaxies: distances and redshifts; galaxies: luminosity function, mass function; galaxies:dwarf; galaxies: fundamental parameters}

\section {Introduction}

The faint end of the galaxy luminosity function 
is a fundamental constraint
on theories of galaxy formation. 
All determinations of the low luminosity slope are dramatically 
shallower than
the  predicted mass function of dark matter halos.
Baryonic physics appears to be the key to resolving this
discrepancy. 

Physical processes possibly
relevant to the faint end slope include the gas cooling time (White \& 
Rees 1978), suppression by photoionization of low-mass galaxy formation 
(Benson et al. 2002), ``feedback'' mechanisms (Benson et al. 2003),
merging and tidal stripping.
Benson et al (2002; 2003) show that various 
combinations of 
these processes lead to very different faint end slopes. 
These differences 
can be a function of galaxy environment.  

Measuring the slope of  the faint end of the galaxy luminosity function, 
$\alpha$, remains
an unresolved observational challenge. For the ``field'' luminosity function   
deeper 
redshift surveys covering 
increasing areas of the sky
provide a route to better and better constraints.
However, in any magnitude limited redshift survey the 
least luminous galaxies occupy a small volume. Because luminosity and surface
brightness are strongly correlated, neither the detection nor
the spectroscopy of the lowest luminosity galaxies is trivial. Thus 
incompleteness at the faint end is a frustrating and serious issue.

Disney \& Phillipps (1983) and McGaugh (1996) emphasized the systematic biases resulting from failure to include low surface brightness galaxies in the determination of the luminosity function. 
Sprayberry et al. (1997) made an early measurement of the impact of low surface brightness galaxies on the
determination of the field luminosity function. Their analysis, based on a catalog of low surface brightness galaxies derived from
APM (Automatic Plate Measuring) scans (Impey et al. 1996), demonstrates that the inclusion of low surface brightness
galaxies substantially steepens the field luminosity function. They obtained a faint end slope
$\alpha = -1.46$ in the B-band. For active star-froming galaxies in the 2dF redshift survey based on 
based on b$_J$ photometry, Madgwick et al. (2002) obtained a faint end slope of $\alpha = -1.5$

Also in the B-band, Driver et al. (2005) analyze the carefully constructed Millenium Galaxy Catalog and derive a faint end slope for the global luminosity function of $\alpha = -1.13\pm 0.02$. They examine the relationship between luminosity and surface brightness for their sample and conclude that the surface brightness distribution is broader for less luminous objects. Driver et al. (2005) show that the faint end slope of the luminosity function is sensitive to the limiting surface brightness of the survey. Earlier B-band work by
Cross \& Driver (2002) had yielded an even shallower faint end slope and indicated robustness to surface brightness issues.

Blanton et al (2005b) made a major step toward measuring and understanding the
behavior of the field luminosity function at low luminosity across all of the SDSS photometric bands. They construct and 
analyze a low redshift sample of galaxies from the Sloan Digital Sky
Survey (SDSS hereafter) to place constraints on $\alpha$. 
Blanton et al (2005b) caution that their sample is not necessarily complete
at least in part because the SDSS was not optimized for this application and they carefully simulate their surface brightness completeness.
They measure $\alpha \sim -1.3$ in the $r$-band. 
Interestingly, this result is similar to the faint end slope of spectroscopically determined
cluster and group luminosity functions extending to comparably low luminosities (e.g. Mahdavi et al. 2005; Rines \& Geller 2010). They also argue that missing low
surface brightness galaxies could steepen the slope to $\alpha \sim -1.5$.
Like Sprayberry et al. (1997), Blanton et al. (2005b) conclude that a majority of the faint galaxy population is blue, low concentration, and low surface brightness.

Here we use a deep complete redshift survey with R$\leq$ 20.6, 
the Smithsonian Hectospec Lensing Survey (SHELS hereafter; Geller et al. 2005; Geller et al. 2010) covering 4 
square degrees of the sky to  constrain the value of $\alpha$. 
The galaxy catalog derives from the Deep Lens Survey (DLS hereafter; Wittman et al. 2006) which reaches
a $1 \sigma$ surface brightness limit of $\mu_R = 28.7$ magnitudes per 
square arcsecond.  We constrain $\alpha$
for redshifts $z \leq 0.1$ focusing on the contribution of blue, low surface brightness galaxies.

We review the SHELS survey in Section 2. Section 2 contains a discussion of 
the survey completeness as a function of surface brightness.  We discuss the
magnitude - surface brightness relation for the survey.
In Section 3, we display some of the lowest luminosity objects in our sample and we derive the luminosity 
function for the sample segregated by surface brightness. In Section 4 we compare the faint-end slope of our
luminosity function with the COSMOS faint-end slope based on photometric redshifts (Liu et al. 2008).
We conclude in Section 5.

\section {The Data}

We use two ambitious surveys to explore the faint end of the galaxy luminosity function.
Generally low luminosity galaxies have low surface brightness. We take advantage of the properties of the two surveys that particularly enable access to these low surface brightness objects.

The DLS (Wittman et al. 2006) is an NOAO key program covering 20 square degrees
in five separate fields; we use the four square degree F2 field at 
$\alpha$ = 09$^h$19$^m$32.4$^s$ and 
$\delta$ = +30$^{\circ}$00$^{\prime}$00$^{\prime\prime}$. The DLS photometric data 
were acquired in a 5-hour integration on the Mayall 4-meter in $<0.9^{\prime\prime}$ seeing and reaching a 1 $\sigma$  limit for source detection in R of 28.7 magnitudes arcsec$^{-2}$ is a good basis for identifying low surface brightness galaxy candidates. We describe our approach to this issue in Section \ref{photometry}.

SHELS (Geller et al. 2005; Geller et al. 2010) is a
redshift survey covering the F2 field to a limiting apparent magnitude 
R = 20.6. SHELS is 98\% complete to R = 20.3, 96\% complete to R = 20.6.
SHELS contains 541 galaxies with R $\leq 20.6$ at z $\leq 0.1$ where 
we can examine the  behavior of the luminosity functions for the lowest
luminosity galaxies. We made a concerted attempt to measure a redshift for each of the lowest surface brightness candidates in the photometric catalog. We describe the completeness of the entire SHELS redshift survey as a function of surface brightness in Section \ref{spectroscopy}.

SHELS covers a small field with deep photometry and spectroscopy. The SDSS covers a very wide field to a much brighter limiting apparent magnitude. We contrast the two surveys in Section \ref{SHELSSDSS}.

\subsection {Photometry}
\label{photometry}

Photometric observations of F2 were made with the 
MOSAIC I imager (Muller et al. 1998) on the KPNO Mayall 4m telescope between 
November 1999 and November 2004. The DLS selected all fields including F2 to 
exclude apparently bright nearby galaxies and to avoid known rich clusters with
redshift $z \lesssim 0.1$. Even though rich clusters are
rare, this selection  biases the density 
at $z \lesssim 0.1$ toward values below the average for the local 
universe as a whole. We revisit this issue in Section \ref{discussion}.  

The R band exposures are the basis for the galaxy catalog in F2. The
effective exposure is 14,500 seconds. Wittman et al (2006) describe the 
imaging reduction pipeline.
The 1$\sigma$ limit for source detection in R is 28.7 magnitudes arcsec$^{-2}$.
To this limit there are 45 sources per square arcminute. The automatic object identification algorithm
produces a complete catalog of objects with surface brightness $\mu_{50,R} \leq 27.0$ within the half-light radius. 

We constructed a galaxy catalog from the R-band source list; we base our luminosity function computation on this R-band catalog.
We selected galaxy candidates with Kron-Cousins R $\leq $ 20.6 for spectroscopic observation. 
The magnitudes are extrapolated total magnitudes; they are extrapolated from isophotal magnitudes
within the limiting 28.7 magnitudes arcsec$^{-2}$ isophote.

Nearly all of the galaxy candidates with R$\leq 20.6$ from the DLS also have SDSS photometry; there are only 104 DLS galaxy candidates  without SDSS photometry. Many of these objects are 
not resolved by the SDSS and thus they have a broad range in surface brightness. We do spectroscopy for these objects as for all other galaxy candidates. When we compare the DLS with the SDSS, we estimate r-band magnitudes from the R-band DLS photometry and from our spectroscopy.

Figure \ref{fig:4panel.ps} shows the classification diagram we use to select galaxies from 
the DLS object list. The upper left hand panel shows all of the 302,574 objects in F2 with R $ < 22.5$. We use the difference between the magnitude within our 1.5$^{\prime\prime}$ fiber, R$_{1.5^{\prime\prime}}$, and the total magnitude R  as a discriminant. 
The upper right panel shows (dark dots) all of the SHELS galaxies with redshifts. There are
15,652 redshifts for galaxies with R$ < 21$ and 12,783 for galaxies with R$ < 20.6$ (vertical green line).  

Objects with the smallest  R$_{1.5^{\prime\prime}}-$R  are mostly stars; objects
with larger R$_{1.5^{\prime\prime}}-$R that lie above the dense galaxy locus
generally result from background fluctuations and various artifacts around bright stars and/or diffraction spikes. 

To construct the catalog for SHELS spectroscopy, we examined all of the 33,038 objects with R$  \leq 20.6$ visually to remove the obvious artifacts. The initial catalog of 33,038 objects is complete to $\mu_{50,R} = 27.0$ at the expense of including artifacts. We conservatively included some apparently stellar objects in the observing list, including any of these objects classified as a galaxy
by the SDSS; our
spectroscopy then showed that $\sim{5}$\% of
the spectroscopically observed objects with  R$\gtrsim 19.5$ are, in fact, stars.  

Very few of the objects that lie above the dense locus of SHELS points are candidate galaxies; we included all objects that were not obviously noise or other artifacts in the spectroscopic observing list. The galaxy candidates are all small compared to the largest angular size objects contained in
the catalog. The galaxy candidates are well away from the photometric thresholds in
Figure \ref{fig:4panel.ps},  implying little or no bias against detection of large low surface
brightness objects in the photometric catalog.

\subsection {Spectroscopy}
\label{spectroscopy}

We acquired spectra for the objects with the Hectospec 
(Fabricant et al. 1998, 2005) on the MMT from
April 13, 2004 to April 20, 2007. The Hectospec observation
planning software (Roll et al. 1998) 
enables  efficient acquisition of a magnitude limited sample. We made a 
concerted effort to acquire spectra for the lowest surface brightness objects.

The SHELS spectra cover the wavelength range 3,700 --- 9,100 \AA\ with a resolution of $\sim$6 \AA.\ Exposure times ranged from 0.75 to 2 hours. The two hour exposures are adequate to yield a redshift even for the lowest surface brightness objects.
The lowest surface brightness objects in the survey required the longer 
integrations. We reduced the data with the standard Hectospec pipeline
(Mink et al. 2007) and derived redshifts with RVSAO (Kurtz \& Mink 1998) with
templates constructed for this purpose (Fabricant et al. 2005). Our 1468
unique pairs of  repeat 
observations imply a mean internal error of 56 km s$^{-1}$ for 
absorption-line objects and 21 km s$^{-1}$ for emission-line objects
(see also Fabricant et al.  2005).

For each spectrum we compute the stellar population age indicator, D$_n$4000. This indicator
is the ratio of flux in the 4000-4100\AA\ band to flux in the 3850-3950\AA\ band (Balogh et al. 1999); it is a measure of the strength of the 4000\AA\ break. The rms scatter in our measurement of D$_n$4000
is 0.086. The internal error in D$_n$4000 is 
only 4.5\% based on our 1468 repeat measurements. A comparison of
overlapping spectra with the SDSS yields a median ratio of 1.00 (Fabricant et al. 2008).
Following Woods et al. (2010) and Kauffmann et al. (2003), we use this indicator to segregate galaxies dominated by old and young stellar populations.

SHELS includes 13,362 galaxies to the limiting apparent magnitude, R = 20.6. 
The integral completeness of the redshift survey to this limit is 96\%.
Geometric constraints are responsible for the  579 objects 
without redshifts; they are mostly near the survey 
corners and edges. On average, Hectospec positionings revisit every region within the DLS field (except for the corners and edges) more than a dozen times. Thus we are minimally biased against close pairs and satellite galaxies.    

\begin{figure}
\centerline{\includegraphics[width=7.0in]{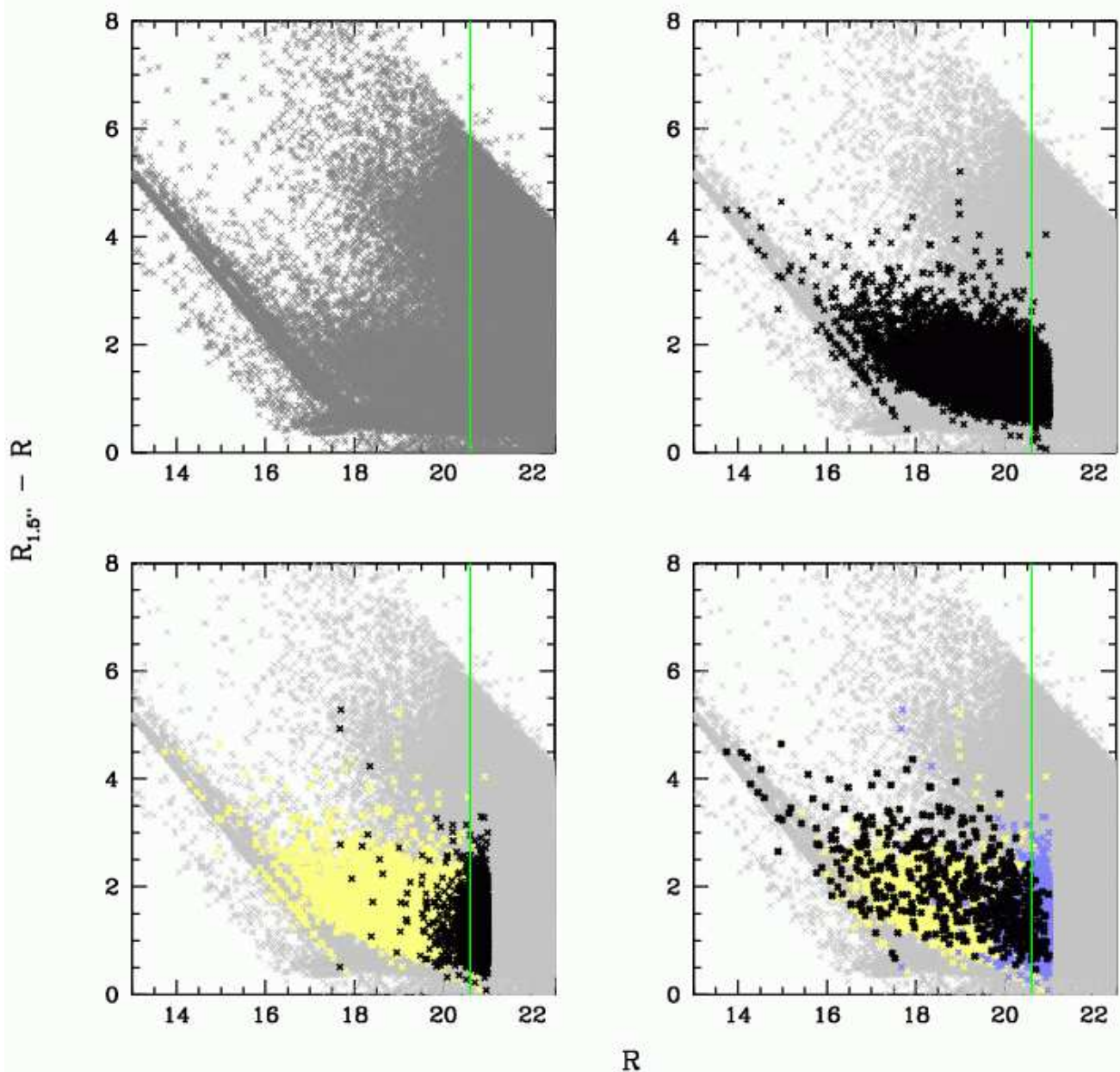}}
\vskip -5ex
\caption{SHELS galaxy selection from DLS photometry. R$_{1.5^{\prime\prime}}$ is the
DLS magnitude within the 1.5$^{\prime\prime}$ Hectospec fiber aperture. R  is the total DLS apparent magnitude. The vertical green line shows the SHELS survey limit R = 20.6. Gray points show all of the objects detected in the DLS survey; the black points (upper right panel) show all of the galaxies with SHELS spectra and R$ \leq 21.$ Most of the ``objects'' at large R$_{1.5^{\prime\prime}}- {\rm R} $ (above the main locus of objects with SHELS spectra)  and with
R$ < 20.6$ are artifacts.  In the lower two panels, yellow points represent all galaxies with measured redshifts. On the lower left, black points represent all galaxy candidates with R$ \leq 21$ and without a measured redshift. In the lower right, black points represent 
SHELS$_{0.1}$ galaxies with redshift $z < 0.1$, blue points are galaxy candidates without a redshift. 
} 
\label{fig:4panel.ps}
\end{figure}

The upper right-hand panel of Figure \ref{fig:4panel.ps} shows  
R$_{1.5^{\prime\prime}}-$R as a function of total magnitude R for all
objects with redshifts and R$ < 21$ (black points). The lower left-hand panel
shows the objects with redshifts as yellow points and galaxy candidates without redshifts as black points. The concentration of objects without spectra toward faint magnitudes is obvious: there are 4238 objects with R $ < 21$ but only 579 with R$ < 20.6$.

The lower right-hand panel of Figure \ref{fig:4panel.ps} shows the sample we use to study the faint end of the galaxy luminosity function, SHELS$_{0.1}$. SHELS$_{0.1}$ includes all SHELS galaxies with $z < 0.1$. There are 541 objects with R $ < 20.6$ and 482 with R $ < 20.3.$ These galaxies
tend to have larger R$_{1.5^{\prime\prime}}-$R because they are
nearby and hence larger on the sky. Most of the objects in the sample with large 
R$_{1.5^{\prime\prime}}-$R are in this low redshift subsample. 

Table \ref{tbl:data} contains the redshifts and R-band magnitudes for the SHELS$_{0.1}$ sample. The Table includes redshifts we obtained for some galaxies fainter than the survey limit R = 20.6 but with redshift z $\leq 0.1$. The Table lists the SHELS identification (column 1), the SDSS objID (column 2), the right ascension (column 3), the declination (column 4), the R-band total magnitude (column 5), the source for the magnitude (column 6), the redshift (column 7), the error in the redshift in km s$^{-1}$ (column 8) , and the observed mean R-band  surface brightness within the half-light radius, $\mu_{50,R}$ (column 9). Among these redshifts, 18 are from the SDSS; the rest are new Hectospec measurements.

\begin{figure}
\centerline{\includegraphics[width=7.0in]{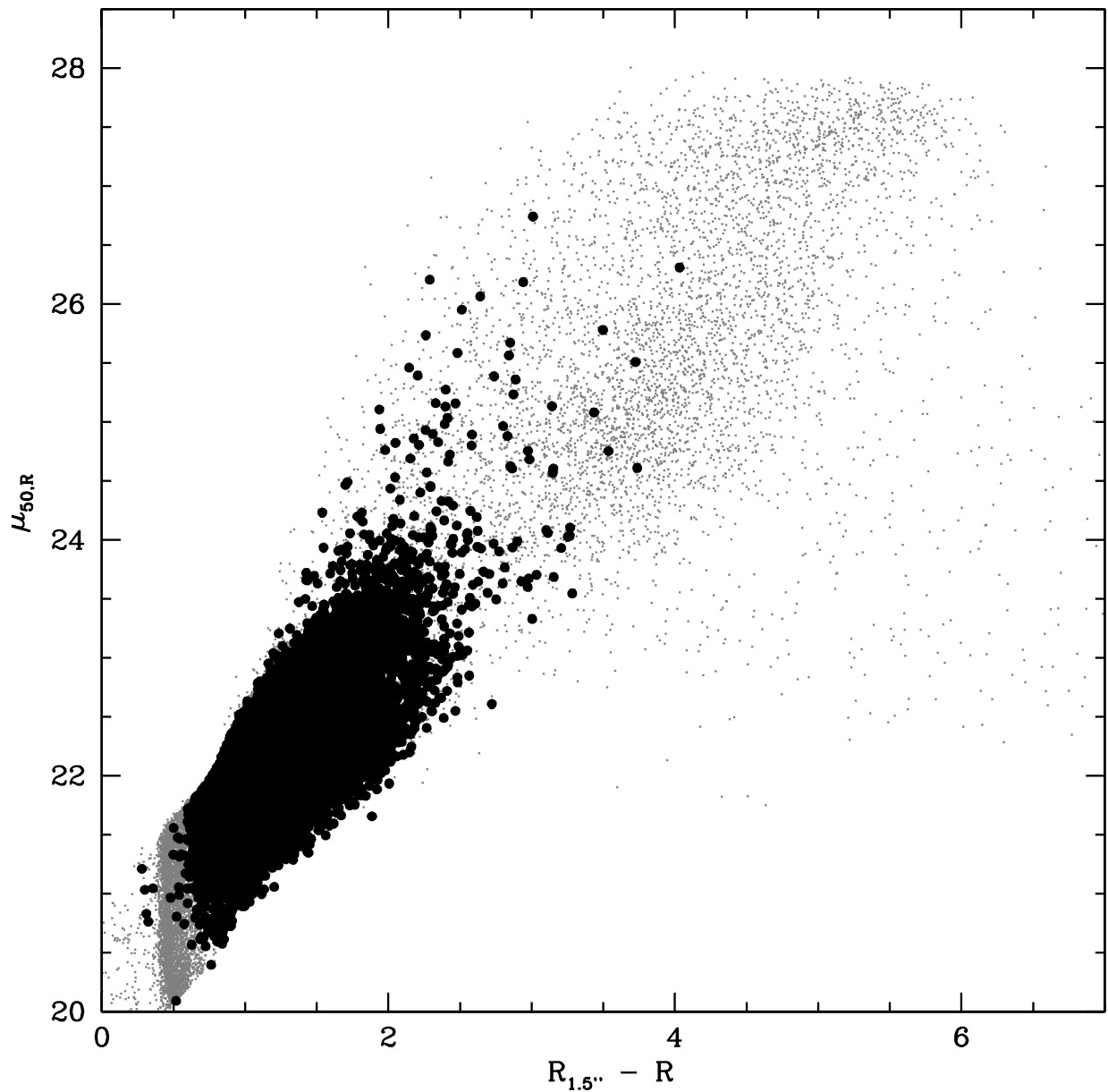}}
\vskip -5ex
\caption{DLS observed mean surface brightness within the half-light radius, $\mu_{50,R}$ as a function
of the magnitude difference R$_{1.5^{\prime\prime}}- {\rm R}$ in the classification diagram of Figure \ref{fig:4panel.ps}. Gray points denote all of the DLS objects in the apparent magnitude range $19 < {\rm R} < 20.6$. Black points denote all of the galaxy candidates
(with or without a redshift). Note that the lowest surface brightness galaxies are all above the survey limit and that most of the low surface brightness objects are artifacts.
} 
\label{fig:SBvMdiff.ps}
\end{figure}

Figure \ref{fig:SBvMdiff.ps} shows the observed mean surface brightness within
the half-light radius, $\mu_{50,R}$ as a function of the discriminant we use to
construct the galaxy catalog, R$_{1.5^{\prime\prime}}-$R  (Figure \ref{fig:4panel.ps}). Gray points indicate all of the objects in the catalog.  We also show (black points) the galaxies in the SHELS redshift survey (including galaxy candidates with or without a redshift and
with $19 < {\rm R} < 20.6$.). We can readily detect objects with $\mu_{50,R} > 24$\ mag arcsec$^{-2}$
and we are complete to $\mu_{50,R} =27$\ mag arcsec$^{-2}$, but galaxy candidates among these objects are rare in this local region of the universe. The lowest surface brightness 
galaxies we find have $\mu_{50,R} \sim 26$ mag arcsec$^{-2}$. It is interesting to note that all of the known dwarfs in the Local Group with M$_R \leq -13$ (the low luminosity limit of our luminosity function; Section \ref{luminosity})  have mean surface brightnesses within the range we can sample (e.g. Mateo 1998; Grebel, Gallagher \& Harbeck 2003). 

\begin{figure}
\centerline{\includegraphics[width=7.0in]{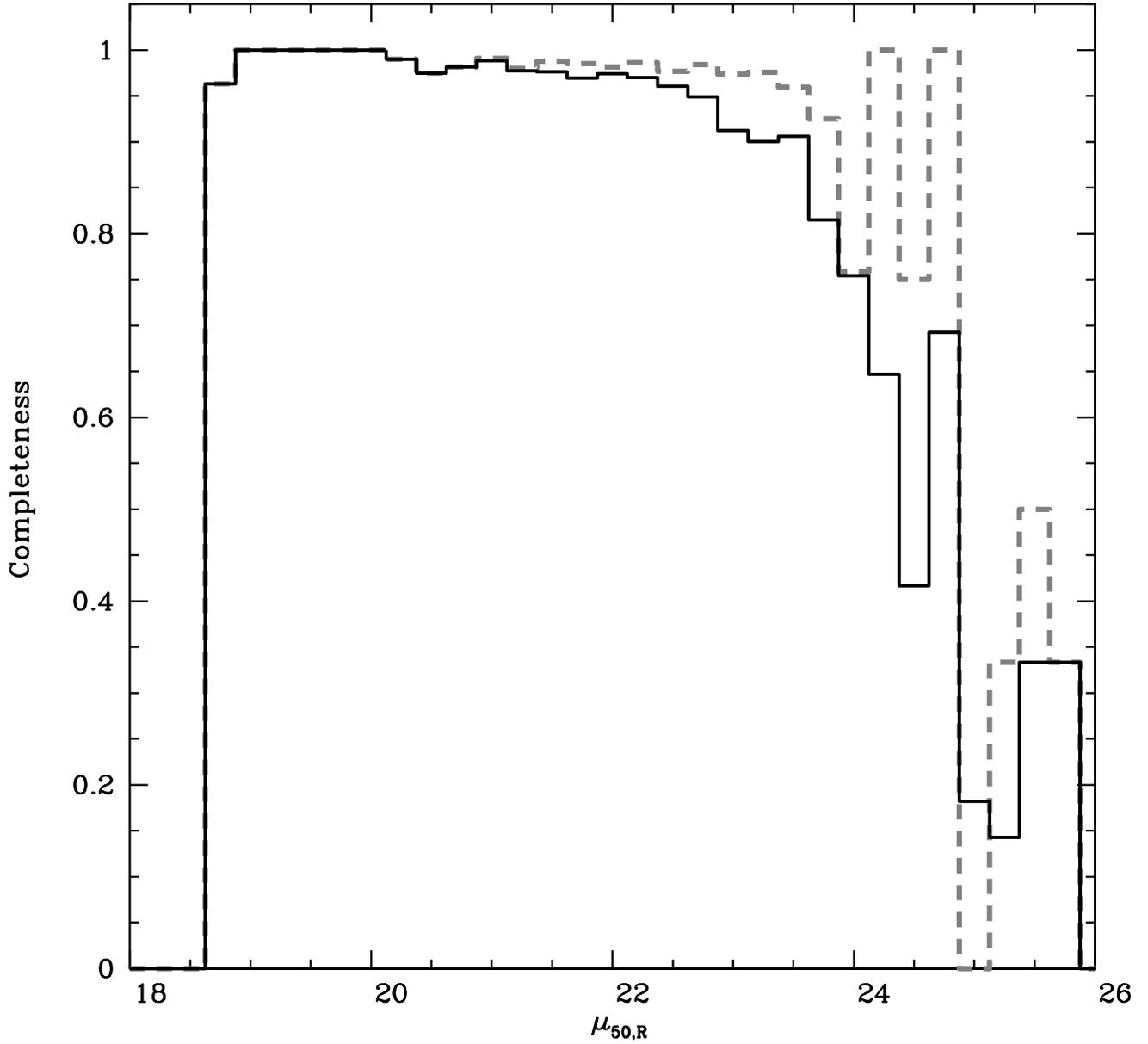}}
\vskip -5ex
\caption{SHELS redshift survey completeness as a function of observed surface brightness,
$\mu_{50,R}$ for galaxies with R$ < 20.3$ (dashed histogram) and for galaxies with R$ < 20.6$
(solid histogram)} 
\label{fig:completeSB.ps}
\end{figure}

Figure \ref{fig:completeSB.ps} shows 
the completeness of the entire SHELS redshift survey as a function of observed mean surface brightness within the  half light radius $\mu_{50,R}$. The completeness is simply the fraction of galaxies in the photometric catalog with measured redshifts as a function of observed 
$\mu_{50,R}$. The dashed histogram shows the fractional completeness for the sample of galaxies with R$ < 20.3$; the solid histogram refers to the sample with R$ < 20.6$. We took substantial care in repeat Hectospec fields to obtain the high level of completeness for R$ < 20.3$. Most of the incompleteness results from  geometric constraints; we fail to sample the corners and edges of the field as well as we sample the central portion.

\subsection {Comparison of SHELS$_{0.1}$  with the SDSS}
\label{SHELSSDSS}

Blanton et al.(2005b) provide a benchmark for measurement of the faint end of the galaxy luminosity function in the local universe in the r-band. They carefully analyze the extensive NYU-VAGC catalog (Blanton et al. 2005a) of SDSS galaxies with $r < 17.77$.  Here we compare the
central average surface brightness distributions and the surface brightness-luminosity relations for the NYU-VAGC and SHELS$_{0.1}$.  Because Blanton et al. (2005) study the r-band luminosity function, we focus on this comparison.

\begin{figure}
\centerline{\includegraphics[width=7.0in]{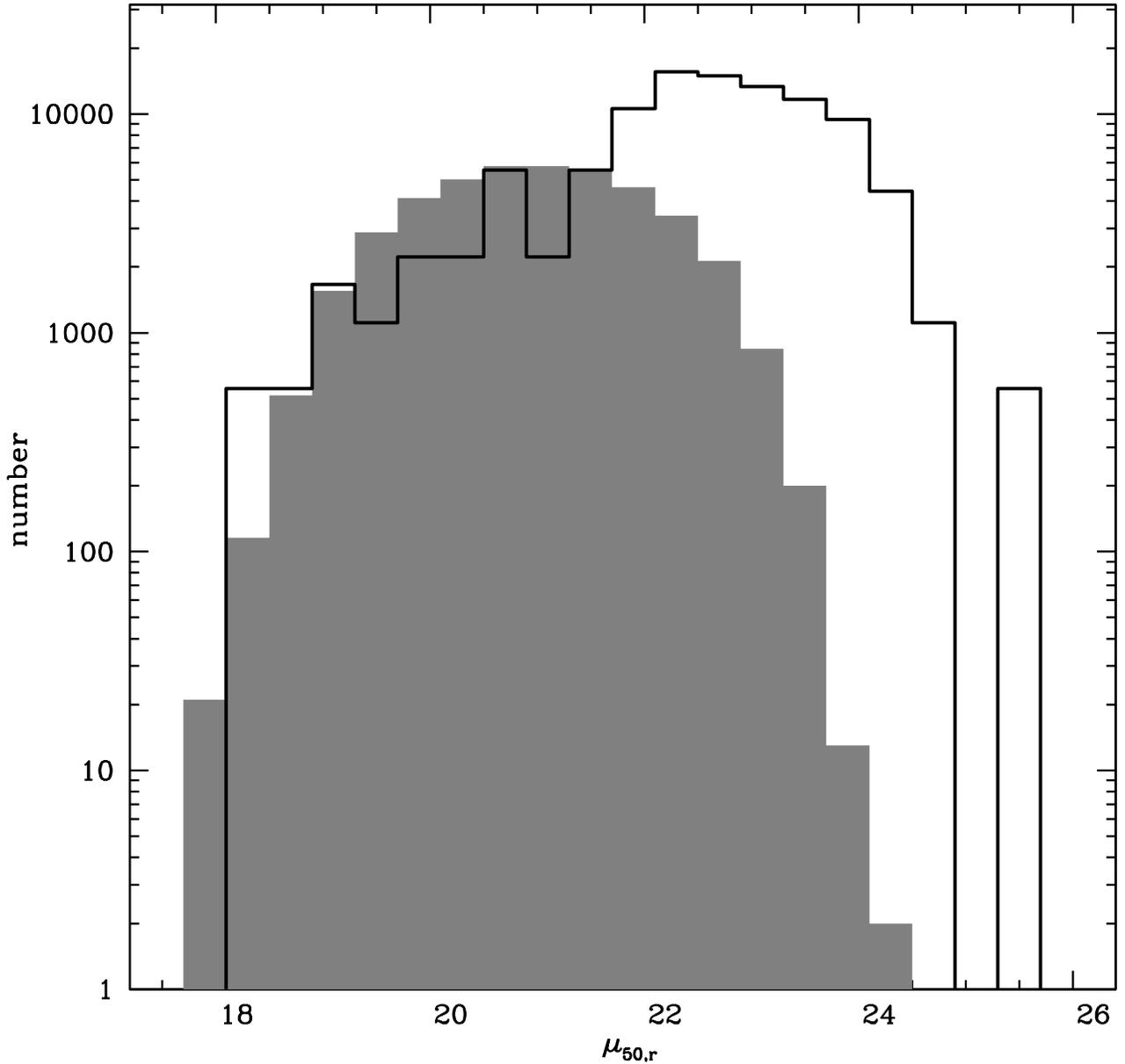}}
\vskip -5ex
\caption{Comparison of the observed surface brightness distribution for the SDSS
sample (solid gray histogram) of Blanton et al. (2005) with galaxies in SHELS$_{0.1}$ that have
$z < 0.05$ (heavy black line). $\mu_{50,r}$ is the
SDSS observed  mean r-band surface brightness within the half-light radius. The vertical axis shows the actual number of SDSS galaxies; we scale the SHELS$_{0.1}$ $z < 0.05$ sample by the relative areal coverage of the two surveys. Note the relatively larger representation of low surface brightness galaxies in the SHELS$_{0.1}$ $z < 0.05$ sample .  
} 
\label{fig:SBcompleteness.ps}
\end{figure}

We first compare the surface brightness distribution of the SHELS$_{0.1}$ sample with the SDSS sample. Figure \ref{fig:SBcompleteness.ps}
shows the distribution of observed mean SDSS $r$-band surface brightness, $\mu_{50,r}$, for the galaxies in the SDSS sample with $z < 0.05$ (solid gray histogram). The heavy solid line 
shows  the  observed $r$-band surface brightness distribution for the SHELS$_{0.1}$ galaxies with $0.02 < z < 0.05$. We scale  the SHELS$_{0.1}$ histogram by the relative area
of the two surveys.  

Figure \ref{fig:SBcompleteness.ps} demonstrates two important aspects of SHELS$_{0.1}$. For observed $\mu_{50,r}$ around the peak SDSS sensitivity, SHELS$_{0.1}$ has systematically fewer objects than the SDSS sample. This deficit in the SHELS$_{0.1}$ survey reflects the 
{\it a priori} selection against nearby galaxies and clusters; the SHELS$_{0.1}$ region is 
underdense by selection.  For observed $\mu_{50,r} \gtrsim 22$, the SHELS$_{0.1}$ survey contains  a relatively much larger number of objects than the SDSS.  This difference results from the SHELS$_{0.1}$ redshift survey fainter magnitude limit and completeness for mostly blue low surface brightness objects. The surface brightness distribution for the NYU-VAGC declines steeply over the surface brightness range 23-24 mag arcsec$^{-2}$; the SHELS$_{0.1}$ distribution over this range is essentially flat. 

\begin{figure}
\centerline{\includegraphics[width=7.0in]{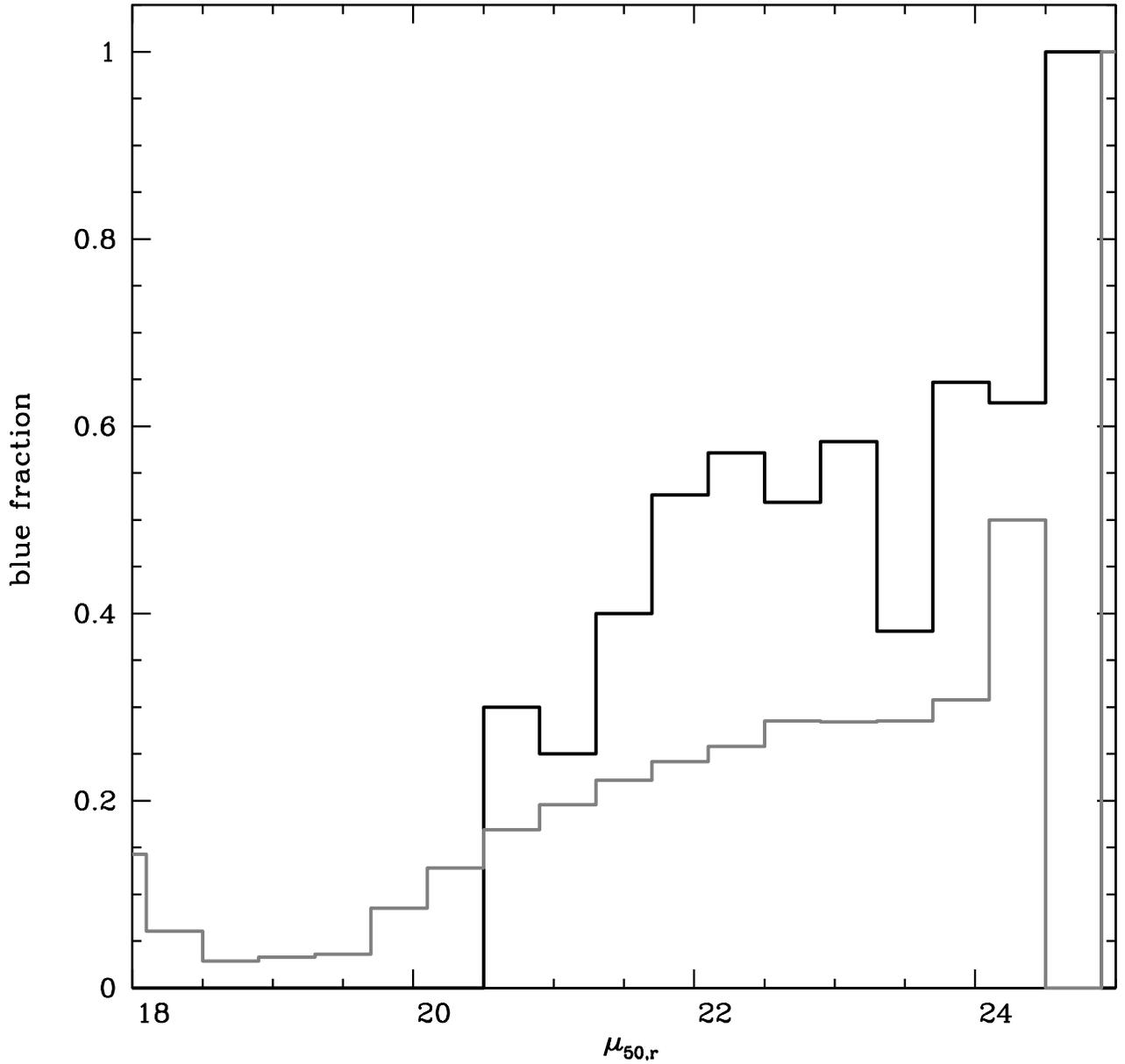}}
\vskip -5ex
\caption{Fractions of galaxies with $g-r < 0.4$ in Figure \ref{fig:SBcompleteness.ps}.
The gray histogram shows the SDSS blue fraction; the black histogram shows the fraction for SHELS$_{0.1}$ galaxies with $0.02 < z < 0.05$. Note the greater blue fraction in the SHELS$_{0.1}$ $z < 0.05$ sample.
} 
\label{fig:bluefraction.ps}
\end{figure}

Figure \ref{fig:bluefraction.ps} shows the fraction of blue objects with $g-r < 0.4$ and with $z < 0.05$  as a function of  $\mu_{50,r}$ for the SDSS (gray histogram) and for SHELS$_{0.1}$; the blue fraction is substantially larger in SHELS$_{0.1}$.

Because the SDSS has a substantial (known) incompleteness for $\mu_{50,r} \gtrsim 23 $, Blanton et al. (2005b) model the incompleteness based on an extrapolation of the relationship
between surface brightness and absolute magnitude for galaxies with M$_r < -18.0$. 
We can examine the relationship between the rest frame surface brightness SB$_{50,R}$ within the half-light radius 
and the absolute magnitude M$_R$ in SHELS$_{0.1}$ and thus test the Blanton et al. (2005b) relationship over a larger range in absolute magnitude. For comparison with Blanton et al. (2005b), the median 
R-r = 0.279 for SHELS$_{0.1}$ galaxies.

The absolute R-band magnitude is

$$M_R = m _R -5{\rm log}d_L -25.0 -k_R(z)$$

\noindent where $d_L$ is the luminosity distance in Mpc for a Hubble constant H = 100h km s$^{-1}$ Mpc$^{-1}$ and $k_R(z)$ is the R-band k-correction derived according to the procedure in
Westra et al. (2010). We use the Hubble constant normalized to 100 for easy comparison with
Blanton et al. (2005b) and others.

The rest frame surface brightness is

$$SB_{50,R} = \mu_{50,R} -10log_{10}(1+z) -k_R(z).$$
 
\noindent This equation implies that, for example, a typical survey galaxy with $\mu_{50,r}$ = 25.0 (at the low surface brightness limit of Figure \ref{fig:bluefraction.ps}) has SB$_{50,R} \simeq 24.3$
at the z = 0.1 survey limit.

\begin{figure}
\centerline{\includegraphics[width=7.0in]{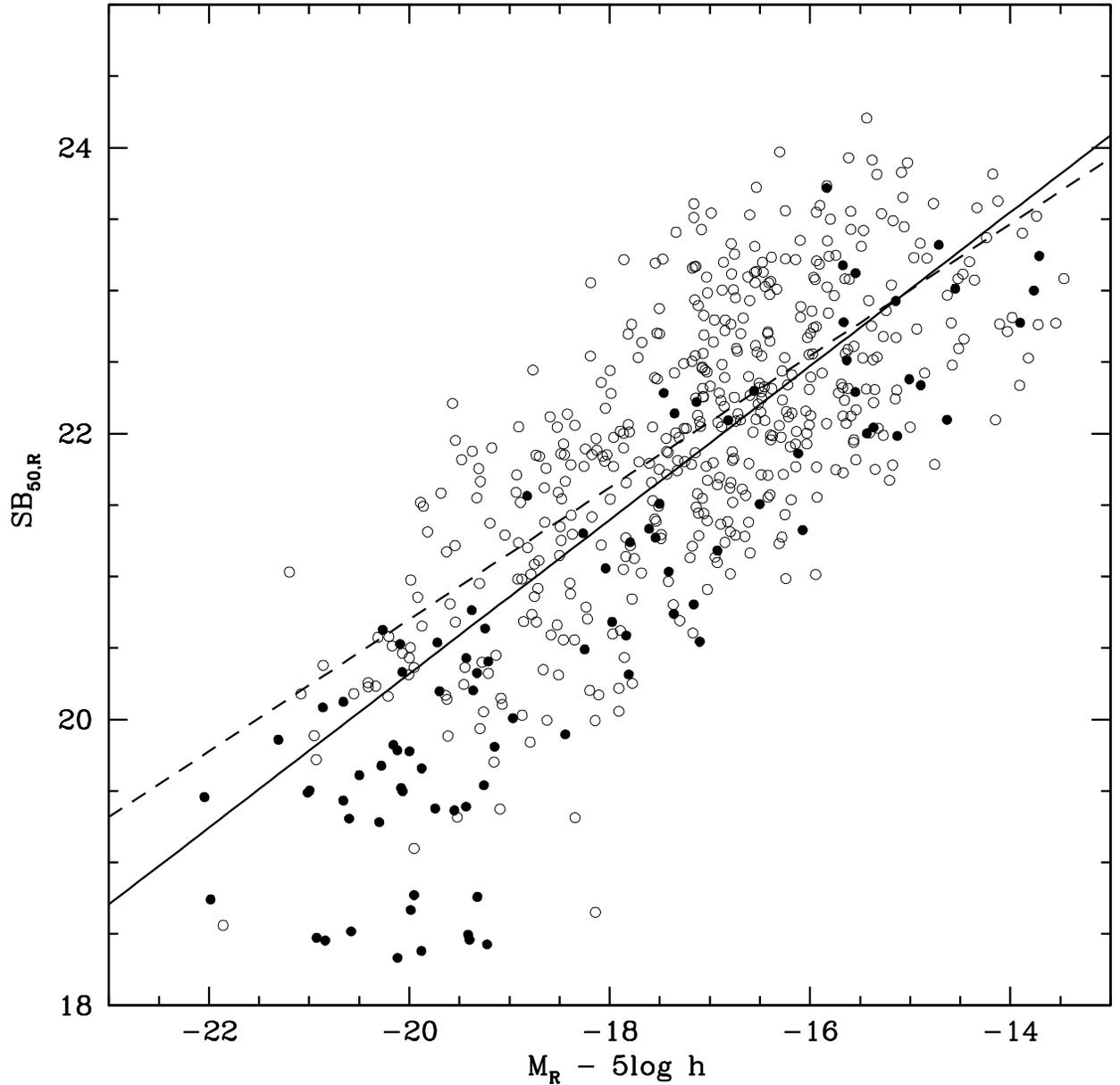}}
\vskip -5ex
\caption{Mean rest frame surface  brightness, SB$_{50,R}$ as a function of absolute magnitude, M$_R$ for SHELS$_{0.1}$. The solid line shows the relation between the two quantities derived from a Bayesian analysis for the full SHELS$_{0.1}$ sample; the dashed line shows the relation for the galaxies with D$_n$4000$ < $ 1.44. Solid dots denote galaxies with D$_n$4000$\geq$1.44;
open circles denote objects with D$_n$4000 $< 1.44$.
} 
\label{fig:SBvAR.ps}
\end{figure}

Figure \ref{fig:SBvAR.ps} shows the relation between SB$_{50,R}$ and absolute magnitude M$_R$. There is only one galaxy in the survey that lands outside the plot limits. It has SB$_{50,R}$ = 25.1 and is obviously contaminated by a nearby bright star; we thus omit it.

We use a Bayesian approach to quantify the correlation between SB$_{50,R}$ and absolute magnitude M$_R$.
This approach is similar to the one used by Blanton et al. (2005b) but not identical.
SB$_{50,R}$ and M$_R$ clearly do not have a one-to-one relation, but, at fixed M$_R$, SB$_{50,R}$ is distributed
according to some probability density distribution (PDF), $p(SB_{50,R}\vert M_R)$. 
In the Bayesian approach, we can infer the parameters of this PDF. We do not need to assume that the spread originates
from random measurement errors around the
{\it ideal} relation $SB_{50,R}=a+bM_R$, as in a standard linear regression. In addition, 
unlike the usual fitting technique, we can model 
the uncertainties in the individual measures as random variates. 

We assume that $SB_{50,R}$ is normally distributed around the mean 
$\langle SB_{50,R}\rangle = a + b M_R$
with variance $\sigma_{\rm int}^2$. We then need to determine the three
parameters $a$, $b$, and $\sigma_{\rm int}$ and their PDFs. 
We assume flat priors for both $a$ and $b$. For the inverse of the 
variances of the individual measures, $1/{\sigma_{SB_{50,R}}}^2$ and $1/{\sigma_{M_R}}^2$, 
we adopt the usual assumption that they are random variates drawn from
a Gamma distribution with large variance (e.g. Andreon and Hurn 2010). 
This choice is appropriate for quantities that are positively 
defined and provides basically flat priors for the uncertainties.

We use the free software JAGS developed by Martyn
Plummer\footnote{www-fis.iarc.fr/$\sim$martyn/software/jags/} to run
Markov Chain Monte Carlo simulations. We estimate the PDFs
of our three parameters by running $3\times 10^5$ iterations.
For the full sample SHELS$_{0.1}$ we find $a=31.08^{+0.36}_{-0.37}$,
$b=0.538^{+0.020}_{-0.021}$, $\sigma_{\rm int}=0.791^{+0.025}_{-0.025}$. Figure
\ref {fig:SBvAR.ps} shows the sample and the result of the Bayesian analysis of the 
entire sample (solid line).

To compare more closely with the analysis of Blanton et al. (2005b), we examine the
SB$_{50,R}$-M$_R$ relation omitting the redder objects. 
Rather than explicitly using color 
to segregate the galaxy populations, we use the redshift independent spectroscopic indicator D$_n$4000. Woods et al. (2010) and Kauffman et al. (2003) show that the distribution of this indicator is bimodal and that it can be used to segregate galaxies with younger stellar populations (generally blue objects) from those dominated by an old stellar population (generally red objects). At low redshift segregation by D$_n$4000 is essentially equivalent to segregation by $g -r$. Woods et al (2010) divide their sample at the local minimum between
the two peaks, D$_n$4000 = 1.44. 

For the 431 blue objects with D$_n$4000$ < 1.44$,
we find $a=29.92^{+0.40}_{-0.39}$,
$b=0.461^{+0.023}_{-0.023}$, $\sigma_{\rm int}=0.742^{+0.026}_{-0.025}$. Figure \ref{fig:SBvAR.ps} shows the 431 objects as open circles, the solid points represent galaxies with D$_n$4000$\geq 1.44$, and the dashed line represents the result of the Bayesian analysis.  The slope $b$ is nearly identical to the value, 0.45, obtained by Blanton et al. (2005b) for SDSS galaxies
with Sersic index $n < 2$ and M$_r < -18$.  

In their model for $p(SB_{50,R}\vert M_R)$, Blanton et al. (2005b) allow for an increase in
$\sigma_{\rm int}$ for less luminous galaxies (for results in the B-band see e.g.
de Jong \& Lacey (2000); Cross \& Driver (2002); and Driver et al. (2005)). Our data do not support such an increase.
For galaxies with M$_R < -18$ we find $\sigma_{\rm int} = 0.766^{+0.050}_{-0.045}$, very similar to the result for the sample as a whole. The value of $\sigma_{\rm int}$ is, however, consistent with the SDSS value at M$_r \sim -18$. The overall consistency of the
SDSS results with the DLS is remarkable.

\subsection {SHELS$_{0.1}$ Redshift Survey Completeness}

\begin{figure}
\centerline{\includegraphics[width=7.0in]{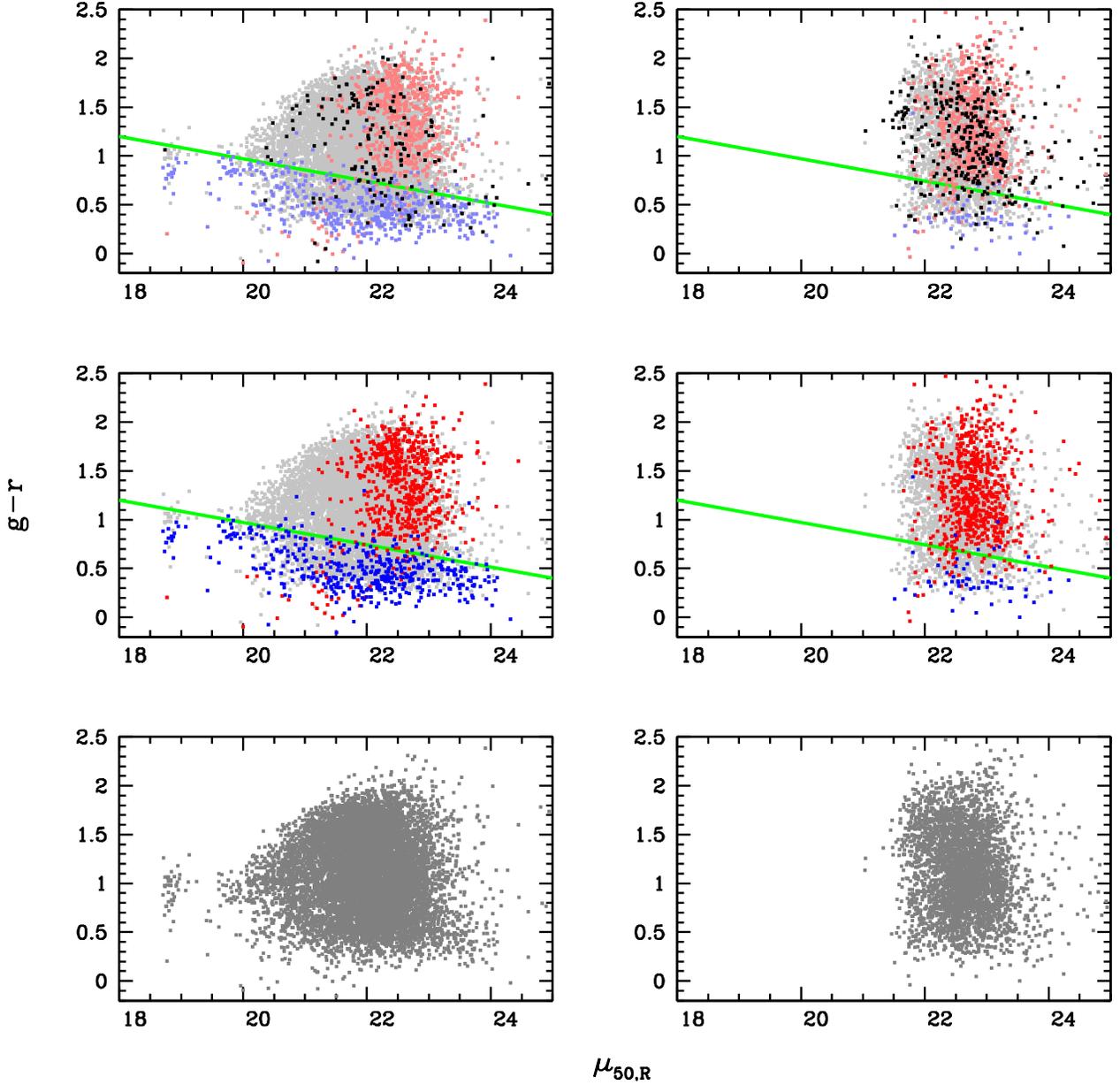}}
\vskip -5ex
\caption{SDSS $g-r$ color as a function of DLS observed surface brightness, $\mu_{50,R}$
 for SHELS. This plot is the basis for an estimate of the completeness of SHELS$_{0.1}$.
The left column applies to galaxies with R$ < 20.3.$; the right-hand column applies to galaxies with $20.3 \leq {\rm R} < 20.6$. The bottom panels show all galaxies in the
magnitude range (gray points). The central panels show the low and high redshift portions
of the sample: blue points represent galaxies with $z < 0.1$ and red points represent galaxies with $z >0.5$. Not surprisingly the galaxies in the $20.3 \leq {\rm R} < 20.6$ interval are mostly at $z > 0.1$ (gray and red points in the upper panel). Most of the low redshift sample lies below the green line. Black points in the upper panels show all of the objects in the photometric survey without a redshift; only a few points overlap the color-surface brightness range spanned by the low redshift sample indicated in blue.
} 
\label{fig:6panel.ps}
\end{figure}

The completeness of SHELS$_{0.1}$ to the limiting apparent magnitude may, in principle,  differ from the survey as a whole. 
Figure \ref{fig:6panel.ps} shows our approach to estimating the completeness of
the SHELS$_{0.1}$ sample. We use a combination of color and surface brightness as a proxy for redshift to evaluate the completeness for SHELS$_{0.1}$ (Kurtz et al. 2007).

The three
panels in the left-hand column of Figure \ref{fig:6panel.ps} refer to the entire SHELS sample with R$ < 20.3$; the right-hand panels show the SHELS sample with $20.3 \leq {\rm R} < 20.6$. All of the panels show
the observed SDSS  $g-r$ color as a function of the mean DLS surface brightness within the half-light radius, $\mu_{50,R}$. The gray points in the bottom panels represent the entire photometric SHELS sample including galaxies with and without a redshift; the black points in the upper panels show the total number of these galaxies that lack a redshift. The red points in the central panel indicate galaxies with redshift $z > 0.5$ and the blue points indicate galaxies with $z < 0.1$, the SHELS$_{0.1}$ sample. Not surprisingly,  galaxies with 20.3$\leq {\rm R} < 20.6$ overlap substantially with the $z > 0.5$ portion of the redshift survey.
 
The green line in the central and upper panels is an arbitrary delineation of the boundary of
the SHELS$_{0.1}$ sample in magnitude-color space.
93\% of the apparently brighter  SHELS$_{0.1}$ galaxies with R$ < 20.3$ with redshifts are blueward of the line and 87\% of the fainter galaxies with 20.3 $ \leq {\rm R} < 20.6$ are blueward of the line. We  use this
admittedly arbitrary line to 
estimate the incompleteness of SHELS$_{0.1}$. The black points in the upper panels show all of the SHELS galaxies without redshifts. To estimate the completeness of SHELS$_{0.1}$ we estimate the fraction of these black points that are probably galaxies with $ z < 0.1$. Note that most of the black points, regardless of apparent magnitude correspond to red galaxies, most probably at high redshift.

Table \ref{tbl:candidates} lists the galaxy candidates below the green line and without redshifts;
these are the objects most likely to be missing from our low redshift sample.
For R$< 20.3$, 21\% of the galaxies with redshifts and below the green line are at $z \leq 0.1$;
for 20.3$\geq {\rm R} < 20.6$ this fraction is 11\%. Note that the gray points below the green line in the middle and upper panels have redshifts between 0.1 and 0.5. Only the black points lack a redshift and statistically we expect that most of these are at $z > 0.1$. In fact, assuming that the fraction with $z \leq 0.1$ is the same among the galaxies without measured redshifts (black points), we expect that only $\sim 14$ of the
86 objects in the Table are at $z \leq 0.1$. We note that many of the objects in Table
\ref{tbl:candidates} are near the edges of the field (particularly the higher surface brightness
objects). 

To estimate the completeness of SHELS$_{0.1}$, we  compute the fraction of SHELS redshifts with $z < 0.1$ {\it both} above and below the fiducial green line. We note that these estimates are insensitive to the exact position of the green line.  
We then assume that these same fractions of the galaxies without redshifts are probably
at $z < 0.1$. For R$ < 20.3$, SHELS$_{0.1}$ is  98\% complete. The differential completeness in the interval $20.3 < {\rm R} < 20.6$ for SHELS$_{0.1}$ is 92\%, greater than the 89\% for the entire SHELS sample because most of the objects without redshifts are small, faint, and red. Photometric redshifts from SDSS substantiate this analysis. In summary, we estimate that SHELS$_{0.1}$ is 97\% complete to R = 20.6; we are missing only 14$\pm$4 objects.   Because our samples are substantially complete, we make no corrections for incompleteness in our calculation of the luminosity function.

\section {The Galaxy Luminosity Function}
\label{luminosity}

\begin{figure}
\centerline{\includegraphics[width=6.5in]{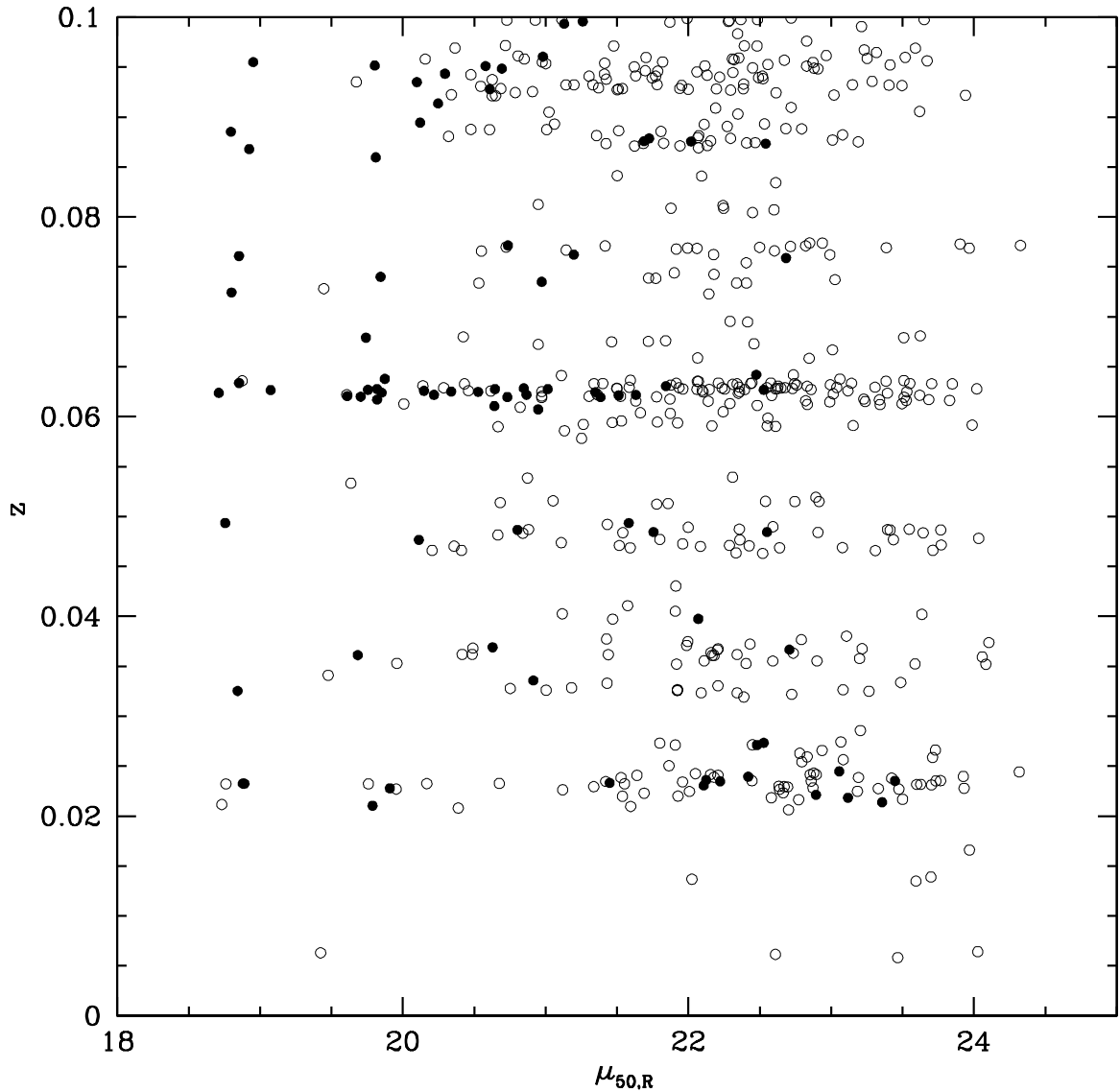}}
\vskip -5ex
\caption{Mean observed surface brightness, $\mu_{50,R}$, as a function of redshift for 
SHELS$_{0.1}$ galaxies. Solid points indicate galaxies with D$_n$4000 $> 1.44$; open circles indicate objects with D$_n$4000 $\leq 1.44$ This division corresponds well with division by color or spectroscopic type. Note that the large D$_n$4000 (redder) galaxies are predominantly in
the densest structure at $z \sim 0.06$.
} 
\label{fig:zVSmu.ps}
\end{figure}

The SHELS$_{0.1}$ survey probes a small region of the universe to a faint, uniform
limiting observed surface brightness. Figure \ref{fig:zVSmu.ps} shows the redshift of each survey galaxy as a function of the observed DLS R-band surface brightness within the half light radius, $\mu_{50,R}$. Large-scale structure in the region is obvious in the highly 
clustered redshift distribution. The distribution of observed surface brightness reaches the survey limit at every redshift. 

Figure \ref{fig:zVSmu.ps} also provides some insight into the galaxy populations. Again, we segregate the galaxy populations based on the spectroscopic indicator D$_n$4000. Open circles indicate  galaxies with a predominantly young population and with 
D$_n$4000$ < 1.44$; the solid circles denote galaxies with 
D$_n$4000$\geq 1.44$. 

In SHELS$_{0.1}$ galaxies dominated by an old population are rare and they tend to be higher surface brightness objects. They appear predominantly in the densest structure in the survey at $z \sim 0.062$, a reflection of the standard morphology-density relation. The general absence of low surface brightness red objects does not result from selection. In fact, the
R-band DLS photometric data are actually more sensitive to these objects than to low surface brightness blue objects.

The lowest surface brightness galaxies are generally also the lowest luminosity objects
(Figure \ref{fig:SBvAR.ps}). Thus Figure \ref{fig:zVSmu.ps} underscores the result previously obtained by Blanton et al.
(2005b): the faint end of the luminosity function is dominated by low surface brightness galaxies dominated by a young stellar population. These galaxies are usually blue. It is interesting that even though we select our galaxies at R, there are no galaxies with predominantly old populations and $\mu_{50,R} \gtrsim 23.5$ \mas. As a result of $(1 + z)^4$ and K-dimming, the lowest surface brightness objects in Figure 2 are at $z > 0.1$ and thus do not appear in Figure 8.

We compute the luminosity function for SHELS$_{0.1}$ with $0.02 \leq z \leq 0.1$ (532 galaxies) and for three subsets of this sample separated by rest-frame
surface brightness within the half-light radius, SB$_{50,R}$. The restriction to $z > 0.02$ reduces the effect of peculiar velocities on the determination of the luminosity function
and sets a low luminosity limit on the luminosity function of M$_{r} = -13.3 + 5 {\rm log}h$.

\begin{figure}
\centerline{\includegraphics[angle=-90,width=6.5in]{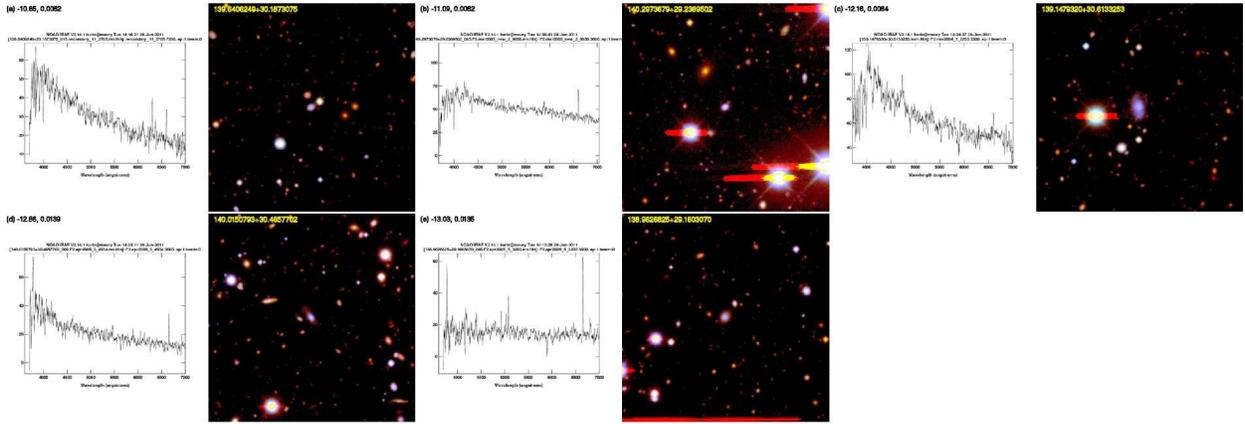}}
\vskip 5ex
\caption{The five lowest luminosity galaxies in SHELS ordered in R-band total luminosity. Galaxy a) has an apparent magnitude R = 20.69, fainter than the SHELS$_{0.1}$ limit. The bold numbers give the absolute R-band luminosity ($h=1$) and the redshift; the yellow numbers in the 3 arcminute square images give the right ascension and declination of the galaxy. Galaxies b)-e) have R $< 20.6$. The objects are all blue and their spectra show Balmer absorption and/or H$\alpha$ emission. 
} 
\label{fig:faintest5.ps}
\end{figure}

The lowest luminosity galaxies in the SHELS$_{0.1}$ are at $z \lesssim 0.015$ and do not enter into the luminosity function calculation. Figure \ref{fig:faintest5.ps} shows the five lowest luminosity galaxies listed in Table \ref{tbl:data}. Galaxies b)-e) all have R$<20.6$ and their
rest frame mean surface brightness within the half light radius  SB$_{50,R} > 22.5$ \mas. Galaxy a), the lowest luminosity galaxy in our sample, is just fainter than the SHELS$_{0.1}$ magnitude limit; it has R = 20.69. Not surprisingly, these galaxies are blue. Many of the spectra show 
Balmer absorption characteristic of a predominantly young stellar population. Most of the spectra show H$\alpha$ emission.

Table \ref{tbl:lf} gives the number of galaxies in each of the samples we analyze. The HSB (high surface brightness) sample includes all galaxies with  
SB$_{50,R} < 21.82$ \mas, the median for the sample. The LSB sample includes the galaxies with
SB$_{50,R} \geq 21.82$ \mas. We note that Blanton et al. (2005b) explore the impact of dividing their sample by surface brightness (Figure 21); they divide their sample at $\mu_{50,r} = 21 \sim SB_{50,R} = 20.7$ \mas, a higher surface brightness by nearly a magnitude arcsec$^{-2}$ than
the median surface brightness for  SHELS$_{0.1}$. 

To explore the dependence on surface brightness more fully we also compute the luminosity function for galaxies with SB$_{50,R} \geq 22.5$ \mas. This subsample, SB$_{50,R}> 22.5$ \mas, satisfies the useful definition of low surface brightness galaxies given by O'Neil (2002): the central surface brightness is a magnitude
fainter than the night sky. To compute our limit we took the 20\% percentile darkest night sky brightness at Gemini, R = 20.4 mag arcsec$^{-2}$
(http://www.gemini.edu/sciops/telescopes-and-sites/observing-condition-constraints/optical-sky-background) as our fiducial value. We model the objects as pure exponential disks to compute our low surface brightness limit, 
SB$_{50,R} > 22.5$ \mas. In this simple exponential disk model, the mean surface brightness within the half-light radius (the quantity we use in this study) is 1.1 magnitudes fainter than the central surface brightness.

We apply the well-known SWML (step-wise maximum likelihood) technique (Efstathiou, Ellis \& Peterson 1988) to compute the luminosity function. We also applied the less widely used (but also non-parametric) C$^-$ (Lynden-Bell 1971) and LCCP (Takeuchi, Ishikawa \& Ichii 2000) techniques to the data; the results are indistinguishable
from the SWML results and for simplicity and clarity, we do not report them here. 

\begin{figure}
\centerline{\includegraphics[width=6.5in]{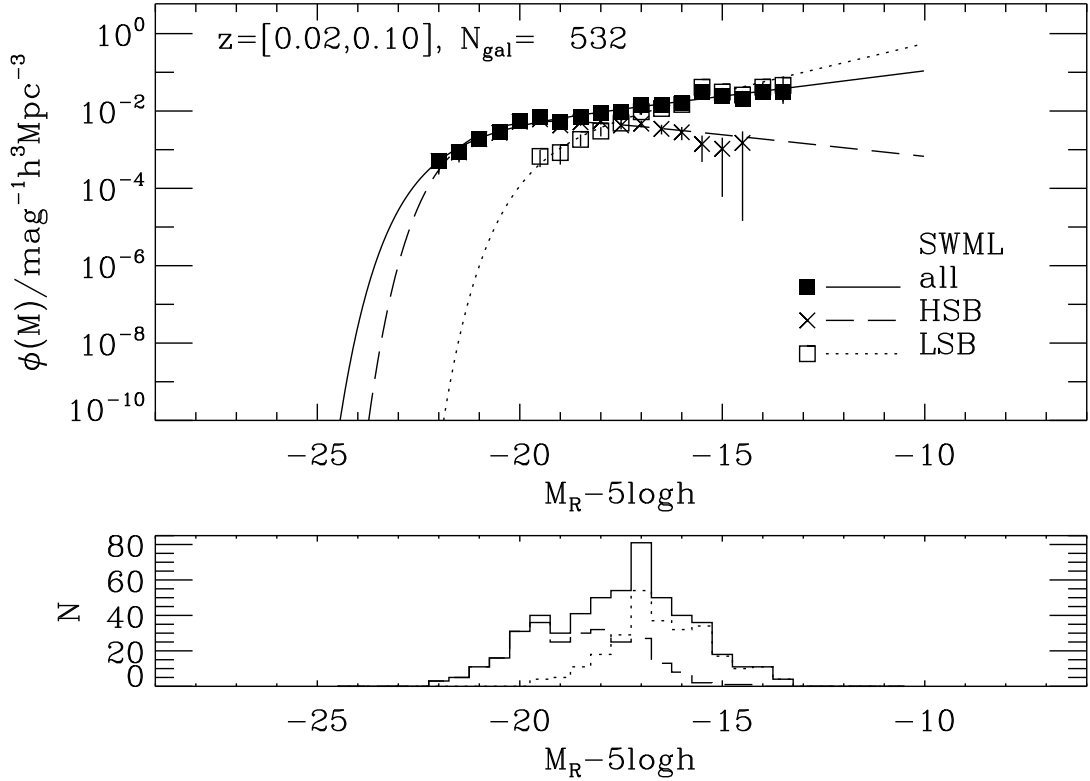}}
\vskip 5ex
\caption{SHELS$_{0.1}$ luminosity functions for all galaxies (solid squares and solid line),
for HSB galaxies with SB$_{50,R} < 21.82$ \mas\ (x's and dashed line), and for LSB galaxies
with SB$_{50,R} \geq 21.82 $ \mas\ (open squares and dotted line). The symbols show the SWML values; the lines are the best fit Schechter function. The histograms show the number of galaxies in each 0.5 magnitude bin for each luminosity function.
} 
\label{fig:lf.SWML.ps}
\end{figure}

\begin{figure}
\centerline{\includegraphics[width=6.5in]{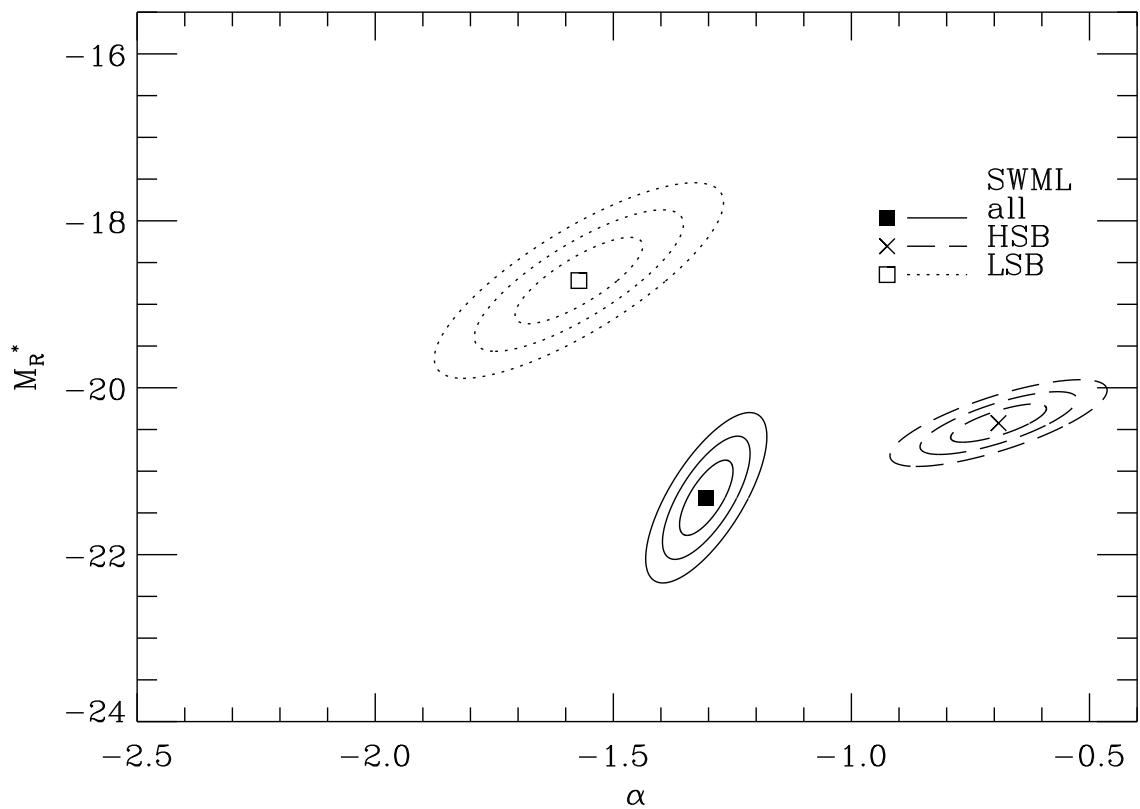}}
\vskip 5ex
\caption{Confidence contours for the SHELS$_{0.1}$ luminosity function parameters for
all galaxies (solid squares and solid line),
for HSB galaxies with SB$_{50,R} < 21.82$ \mas\ (x and dashed line), and for LSB galaxies
with SB$_{50,R} \geq 21.82$ \mas\ (open square and dotted line). The symbols show the best fit Schechter parameters. The contours indicate the 1,2, and 3$\sigma$ limit for the parameters. 
} 
\label{fig:lf.conf.level.SWML.ps}
\end{figure}

\begin{figure}
\centerline{\includegraphics[width=6.5in]{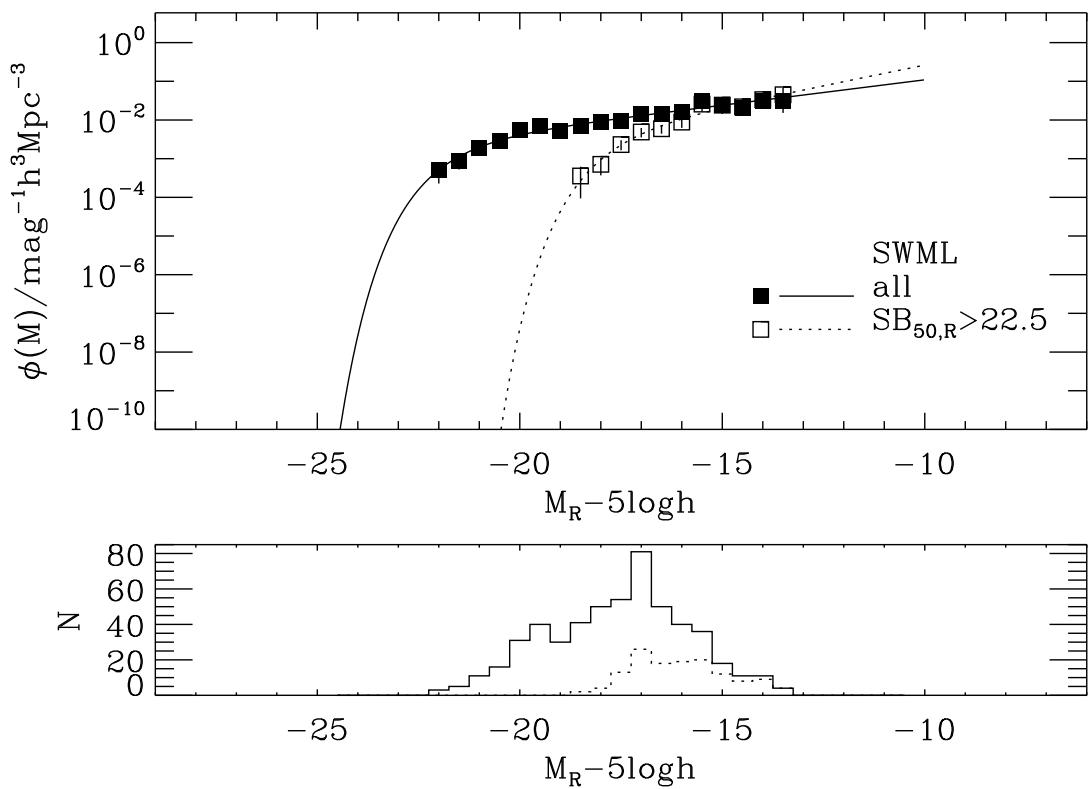}}
\vskip 5ex
\caption{SHELS$_{0.1}$ luminosity functions for the entire SHELS$_{0.1}$ sample (solid squares)
compared with the SB$_{50,R} \geq 22.5$ \mas\ subset (open squares), an unusually low surface brightness sample. The histograms show the number of galaxies in each 0.5 magnitude bin of the
respective luminosity functions.
} 
\label{fig:lf.SWML.LSB.ps}
\end{figure}

\begin{figure}
\centerline{\includegraphics[width=6.5in]{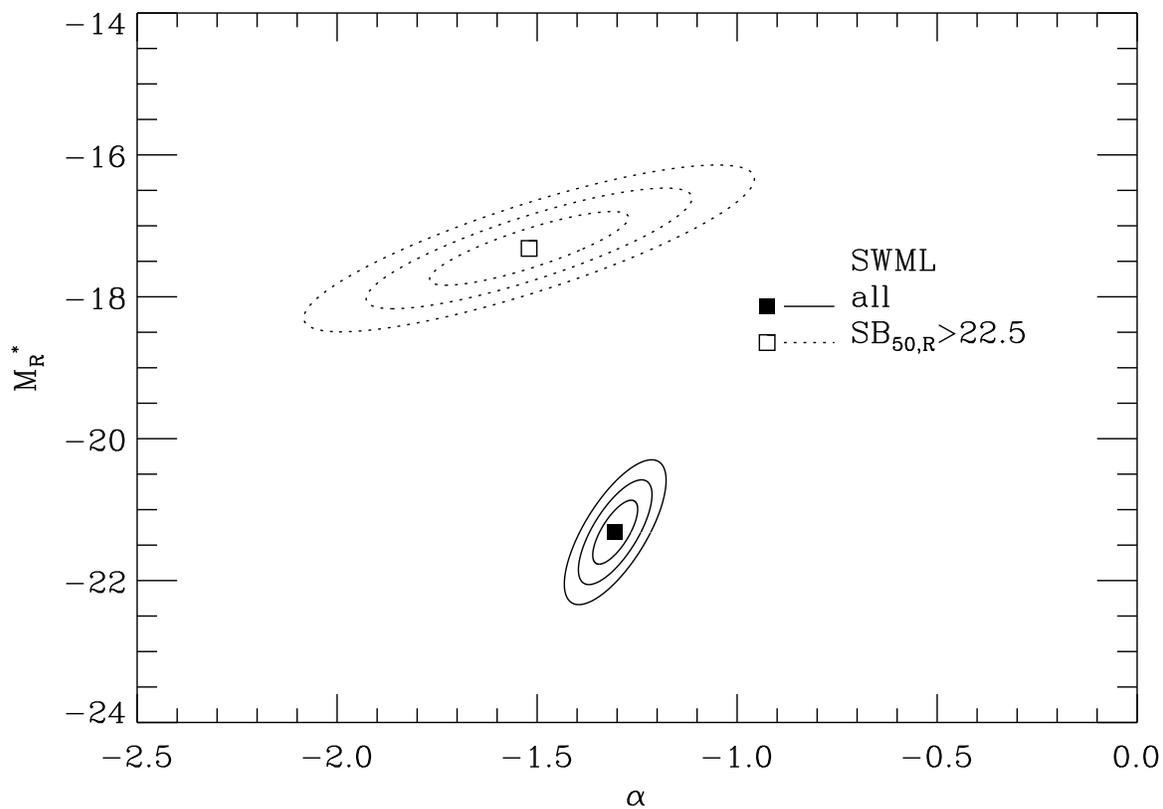}}
\vskip 5ex
\caption{Confidence contours for the sample from Figure \ref{fig:lf.SWML.LSB.ps} with
SB$_{50,R} \geq 22.5$ \mas\ (dotted contours) and for the entire SHELS$_{0.1}$ sample (solid contours). For the low surface brightness sample the error ellipses admit a wide range of values of $\alpha$ and do not exclude $\alpha = -2$. The best fit is $\alpha = -1.52 \pm 0.15$. These low surface brightness galaxies dominate the determination of the faint end slope for the luminosity function of the sample as a whole.
} 
\label{fig:lf-conf-level.LSB.SWML.ps}
\end{figure}

Figures \ref{fig:lf.SWML.ps} and \ref{fig:lf.SWML.LSB.ps}  show the results of the luminosity function calculation for the entire sample and for the three subsamples. Figures
\ref{fig:lf.conf.level.SWML.ps} and \ref{fig:lf-conf-level.LSB.SWML.ps} show the corresponding confidence contours of the luminosity function parameters. The points in the luminosity function plots show the SWML results. We use the bootstrap method to compute the uncertainty at each point. We resample
the galaxy sample 50 times for each luminosity function computation.

Of course, the SWML technique does not assume a form for the luminosity function. We represent the SWML results with a fit to a single Schechter (1976) function:

$$\phi(M)dM = 0.4ln10\phi^*10^{0.4(\alpha + 1)(M^* - M)}exp[-10^{0.4(M^*-M)}]dM$$

\noindent where $\alpha$ is the faint end slope, M$^*$ is the characteristic magnitude, and
$\phi^*$ is the normalization. With their much larger dataset Blanton et al. (2005) fit
a double Shechter function. We obtain reasonable $\chi^2$ for the single Schechter function fits (Table \ref{tbl:lf}).

Table \ref{tbl:lf} lists the luminosity function parameters for the entire SHELS$_{0.1}$ sample and for each of the three subsamples. Table \ref{tbl:lf} also lists the luminosity density for each sample. Application of the parametric STY method (Sandage, Tammann \& Yahil 1979) to our data yields similar results.

Because the volume of SHELS$_{0.1}$ is small, the bright end of the luminosity function
and the values of M$^*_R$ are poorly constrained and cannot be compared with other estimates; the values of M$^*_R$ and of the luminosity density, ${\cal L}$, are only useful for comparison of different subsamples of the SHELS$_{0.1}$ data. Our intent here is to focus on the faint end of the luminosity function and on the contribution of low surface brightness galaxies to the slope.

For the sample as a whole the faint end slope, $\alpha = -1.31 \pm 0.04$, is reasonably well-determined. This result is the same as the r-band faint-end slope  Blanton et al (2005) obtained without correction for missing low surface brightness objects ($\alpha_2 = -1.34 \pm 0.01$). 
This faint end slope is also consistent with other measurements of field and cluster
luminosity functions derived for low redshift samples with complete spectroscopy (e.g. Christlein \& Zabludoff 2003; Mahdavi et al. 2005; Rines \& Geller 2010). 

The luminosity density for the SHELS$_{0.1}$ sample is lower than the value obtained by Blanton et al (2005b) and others as expected based on the selection of the region. We also note that the $\chi^2_\nu$ per degree of freedom is reasonable for the single Schechter function fit; the sample is too small to support the more complex approach of fitting a double Schechter function as in Blanton et al (2005b). 

The luminosity functions for the HSB and LSB subsamples
demonstrate that the lower surface brightness galaxies dominate  the faint end of the luminosity function. The slope for the HSB subsample is quite shallow, $\alpha_{HSB} = -0.69\pm0.07$. The faint-end slope for the LSB sample is $\alpha_{LSB} = -1.57 \pm 0.09$.  This value is 
somewhat steeper than the fit Blanton et al. (2005b) obtain for their r-band luminosity function corrected for missing low surface brightness galaxies ($\alpha_2 = -1.40 \pm 0.01$). 

Because there is little correction for surface brightness incompleteness to our limiting magnitude, the steep faint end slope is an empirical determination of the impact of low surface brightness galaxies on the luminosity function. We make an additional empirical test of this conjecture by extracting a subsample of SHELS$_{0.1}$ that lies within 1$^\circ$ of the field center where the integral completeness of SHELS to R = 20.6 is 97\% (rather than 96\% in the full area). In this region there are 416 galaxies with $z < 0.1$ and we estimate that the
completeness of SHELS$_{0.1}$ is 99\% according to the technique demonstrated in Figure 7.
In other words we are missing 6$\pm$2 galaxies. When we recompute the luminosity functions for all of the subsamples in surface brightness considered here, the faint end slopes agree with those in Table 3 to within 1$\sigma$ for the original sample. 

The value of 
M$^*_R$ for the LSB sample is fainter than for the HSB sample as expected from the correlation between surface brightness and luminosity. It is interesting that the fraction of the luminosity density contributed by the LSB half of the sample  is only 16\% to the limiting surface brightness and absolute magnitude we sample.

Our lowest surface brightness sample SB$_{50,R} > 22.5$ \mas\ contains 135 galaxies and thus the error in $\alpha$ is large. However, there is no other published luminosity function derived from a highly complete redshift survey of such low surface brightness galaxies. The faint end slope is $\alpha_{22.5} = -1.52 \pm 0.16$, consistent with the slope we obtain
for the larger LSB sample. 

\section {Discussion}
\label{discussion}

Determination of the faint end slope of the galaxy luminosity function is sensitive to the
inclusion of low surface brightness galaxies. Although these galaxies make a relatively small contribution to the total luminosity density, they dominate the count of objects at low luminosity. 

The identification of low surface brightness galaxies from a photometric survey is a challenging problem in itself and the acquisition of a spectroscopic redshift for the lowest surface brightness objects
is time-consuming even with a large telescope. One alternative approach is the use of photometric redshifts. Here we compare our results with the luminosity function
obtained by Liu et al. (2008) for the redshift range $z = 0.02 - 0.1$ in the
2 deg$^2$ COSMOS field. In this redshift bin they compute the faint end slope to a limiting
M$_V \sim -12.8$, comparable with our R-band limit of $\sim -13.0$ for the same redshift interval.

Liu et al. (2008) use photometric redshifts computed according to the method of Mobasher et al. (2007). To compute the luminosity function, Liu et al. (2008) treat each galaxy as a weighted probability-smoothed luminosity distribution. They apply a modified version of the 1/V$_{max}$ method and use numerical simulations of their procedure to assess the
biases and random errors in their technique. 

For their entire sample, Liu et al. (2008) obtain a faint-end slope, $\alpha_{COSMOS} = -1.24 \pm 0.07$, remarkably consistent with our $\alpha = -1.31 \pm 0.04$. Liu et al. (2008) segregate their sample according to spectral energy distributions roughly corresponding to various morphological types. Their ScD+Irr bin is probably the most comparable with the low surface brightness portion of our sample. For this subsample, they obtain $\alpha_{COSMOS, ScD + Irr} = -1.46 \pm 0.07$ in essential agreement with our $\alpha_{LSB} = -1.57 \pm 0.09$. 
It is impressive that two very different techniques yield such similar results over the same redshift range.

Liu et al. (2008) comment that the faint-end slope in their $z = 0.02 - 0.1$ bin may be ``qualitatively'' dominated  by low surface brightness dwarfs that are not detected in their survey at higher redshift. The SHELS$_{0.1}$ faint-end slope {\it is} dominated by low surface brightness dwarfs. Thus the essential agreement of the COSMOS low redshift faint-end slope with SHELS$_{0.1}$ argues strongly that  low surface brightness dwarfs (and not evolution) account for the relatively steeper faint end slope at the lowest redshifts in the COSMOS sample. The comparison of SHELS$_{0.1}$ with the COSMOS results underscores the importance of cleanly defined and measured surface brightness limits in surveys addressing the galaxy luminosity function and its evolution.

\section {Conclusion}

Measurement of the faint end slope of the galaxy luminosity function requires attention to lower surface brightness objects that dominate the count at lower luminosities. This issue is, of course, important at all redshifts and failure to understand surface brightness limits may lead to apparent evolution of the faint end of the luminosity function with redshift.  We use the
SHELS redshift survey of one of the DLS survey fields to evaluate the faint end slope at low redshift and to examine its sensitivity to surface brightness.  The DLS photometry offers the possibility of identifying galaxies with lower surface brightness (the photometric survey is complete to a mean surface brightness within the half-light radius of 27.0 mag arcsecond$^{-2}$ at R) and, in carrying out the spectroscopic survey, we paid attention to acquiring redshifts for  these lower surface brightness objects. The lowest surface brightness objects we identify by careful inspection of all of the galaxy candidates in the field are above the photomettric detection limit. Two-hour integrations with Hectospec on the MMT are adequate to acquire a redshift even for the lowest surface brightness objects in the photometric catalog. 

We calculate the faint end slope in the R-band for  the subset of SHELS galaxies with redshifts in the range 0.02$\leq z < 0.1$, SHELS$_{0.1}$. This sample contains 532 galaxies
with R$< 20.6$ with a median surface brightness within the half light radius of SB$_{50,R}$ = 21.82 \mas. We estimate that there are only 14$\pm$ 4 objects missing from this sample. We used this sample to make one of the few direct measurements of  the dependence of the faint end of the galaxy luminosity function on surface brightness.

We compare the  properties of the SHELS$_{0.1}$ galaxies with the much larger, but shallower NYU-VAGC derived from the SDSS (Blanton et al. 2005a, 2005b). We show that  SHELS$_{0.1}$ has a fainter median 
observed surface brightness $\mu_{50,r}$ within the half light radius. The SHELS$_{0.1}$ sample thus enables a strong test of the relation between absolute magnitude and surface brightness
that Blanton et al. (2005b) use to correct their luminosity function for missing low surface brightness objects. We use a Bayesian approach to evaluate the correlation between magnitude and surface brightness; we derive a slope of 0.46$\pm 0.02$, essentially identical to the slope derived by Blanton et al (2005b) for galaxies with M$_r < -18 + 5{\rm log}h$.  Our R-band relation extends to M$_R \sim -14 + 5{\rm log}h$ with essentially constant variance around the mean relation.  

We compute the faint end slope of the luminosity function for the SHELS$_{0.1}$ sample as a whole and for three subsamples segregated by mean surface brightness within the half light radius. For the sample as a whole the faint end slope $\alpha = -1.31 \pm 0.04$, is consistent with both the Blanton et al. (2005b) analysis of the SDSS and the Liu et al. (2008) analysis
of the COSMOS field. This consistency is impressive given the very different approaches of these three surveys: SDSS is a large area shallow survey with spectroscopy; COSMOS is a deep photometric survey with an analysis based on photometric redshifts;  and SHELS$_{0.1}$ is a
dense spectroscopic survey with a 6.5-meter telescope of a photometric catalog derived from 5 hours of integration in better than 0.9$^{\prime\prime}$ seeing on a 4-meter telescope.

A magnitude limited sample of 135 galaxies with optical spectroscopic redshifts and with mean surface brightness, SB$_{50,R} \geq 22.5$
\mas\ is unique to SHELS$_{0.1}$. The faint end slope is $\alpha_{22.5} = -1.52\pm 0.16$, consistent with previous limits on similarly low surface brightness populations from independent samples. Because SHELS$_{0.1}$ samples a low density region of the universe by construction, these low surface brightness objects are predominantly blue. Surveying a larger volume to the depth of SHELS$_{0.1}$ would be an important basis for 
exploring the dependence of the faint end of the luminosity function on environment and galaxy type.  

\acknowledgements

We thank Scott Kenyon for many insightful discussions. We thank Beth Willman for a discussion about Local Group dwarfs. We also thank Michael Strauss and Marijn Franx for encouraging us to use the SHELS survey as a basis for studying the local luminosity function. We thank Michael Blanton and his collaborators for providing the NYU-VAGC on line
and we thank Martyn Plummer for his publicly available JAGS code.
Perry Berlind and Michael Calkins operated the
Hectospec efficiently and with extraordinary expertise. We thank Susan Tokarz for her dedication to careful data reduction. Nelson Caldwell deftly
manages Hectospec queue scheduling for optimal scientific results. We thank the anonymous referee for urging us to improve the clarity of the presentation here. This paper makes use of data from the Sloan Digital Sky Survey. The Smithsonian Institution
partially supported this research. NSF grant AST-0708433 supports Ian Dell'Antonio's research at Brown University. INFN grant PD51 and the PRIN-MIUR-2008 grant \verb"2008NR3EBK_003" ``Matter-antimatter asymmetry, dark matter and dark energy in the LHC era'' partially support Antonaldo Diaferio.

{\it Facilities:}\facility {MMT(Hectospec)}

\clearpage
\begin{landscape}
\begin{deluxetable}{lccccccccc}
\tablecolumns{10}
\tablewidth{0pc}
\tabletypesize{\footnotesize}
\tablenum{1}
\tablecaption{SHELS$_{0.1}$ Redshifts\tablenotemark{a}}
\tablehead{ 
   \colhead{SHELS id} &
   \colhead{SDSS objID} &
   \colhead{ra$_{2000}$} &
   \colhead{dec$_{2000}$} &
   \colhead{R} &
   \colhead{R Source\tablenotemark{b}} &
   \colhead{z} &
   \colhead{z Source\tablenotemark{c}} &
   \colhead{$\delta$z} &
   \colhead{$\mu_{50,R}$}\\& & & & & & & & 
   \colhead {km s$^{-1}$}& 
   \colhead {mag arcsec$^{-2}$}
}
\startdata
138.7082030+30.4863253 & 587738947740697041 & 9:14:49.969 & 30:29:10.771 & 19.5084 & DLS & 0.0218412 & MMT & 32.3 & 22.5807 \\
138.7106146+30.1474856 & 588017978876101272 & 9:14:50.547 & 30:08:50.948 & 19.6469 & DLS & 0.0373713 & MMT & 47.1 & 24.1041 \\
138.7245438+30.2541523 & 588017978876166576 & 9:14:53.891 & 30:15:14.948 & 18.8388 & DLS & 0.0236177 & MMT & 32.1 & 22.1238 \\
138.7254512+30.6434571 & 588017979413103093 & 9:14:54.108 & 30:38:36.446 & 18.7908 & DLSm & 0.0238024 & MMT & 81.9 & 23.4231 \\
138.7268120+30.4084400 & 587738947740696966 & 9:14:54.435 & 30:24:30.384 & 20.9377 & DLS & 0.0628633 & MMT & 40.9 & 22.8286 \\
138.7271334+29.2781541 & 588017977802228132 & 9:14:54.512 & 29:16:41.355 & 19.4561 & DLS & 0.0206206 & MMT & 50.6 & 22.7012 \\
138.7289058+30.2024933 & 588017978876166301 & 9:14:54.937 & 30:12:08.976 & 15.7759 & DLS & 0.0232745 & MMT & 18.4 & 20.6754 \\
138.7325010+30.0934152 & 587738947203760418 & 9:14:55.800 & 30:05:36.295 & 18.3344 & DLS & 0.0238612 & MMT & 43.6 & 23.1907 \\
138.7353477+30.4341614 & 587738947740696611 & 9:14:56.483 & 30:26:02.981 & 17.0162 & DLS & 0.0222926 & MMT & 34.3 & 21.69 \\
138.7358290+30.2827235 & 588017978876166629 & 9:14:56.599 & 30:16:57.805 & 20.6605 & DLS & 0.0222285 & MMT & 47.6 & 23.1852 \\
\enddata
\tablenotetext{a}{Table 1 is published in its entirety in the electronic version of the 
{\it Astronomical Journal.} The first ten lines are shown here for guidance regarding its 
form and content.}
\tablenotetext{b} {The R-band magnitude source is either the DLS or the SDSS. We translate SDSS 
$r$ to R using the redshift and the SDSS colors. DLSm or SDSSm indicates that the magnitudes required a detailed calculation outside the pipeline for the survey generally as a result of a nearby star and/or an artifact in the imaging data.}
\tablenotetext{c} {The redshift $z$ is from the Hectospec on the MMT (MMT) or from the SDSS. The
error ${\delta}z$ is also from the redshift source.}
\label{tbl:data} 
\end{deluxetable}
\end{landscape}

\clearpage

\begin{deluxetable}{lccccccccc}
\tablecolumns{7}
\tablewidth{0pc}
\tabletypesize{\footnotesize}
\tablenum{2}
\tablecaption{Low Redshift Candidate List\tablenotemark{a}}
\tablehead{ 
   \colhead{SHELS id} &
   \colhead{SDSS objID} &
   \colhead{ra$_{2000}$} &
   \colhead{dec$_{2000}$} &
   \colhead{R} &
   \colhead{$\mu_{50,R}$} &
   \colhead{$g-r$\tablenotemark{b}}\\& & & & & 
   \colhead{mag arcsec$^{-2}$}&  
}
\startdata
138.7089504+30.9843187 & 587738948277698932 & 9:14:50.148 & 30:59:03.547 & 20.5556 & 23.3638717 & 0.521\\
138.7255980+30.0542613 & 587738947203760545 & 9:14:54.144 & 30:03:15.341 & 20.5392 & 23.4122087 & 0.507\\
138.7264737+30.3206758 & 588017978876166652 & 9:14:54.354 & 30:19:14.433 & 19.1392 & 23.6712299 & 0.265\\
138.7291346+30.4367746 & 587738947740697008 & 9:14:54.992 & 30:26:12.389 & 20.3627 & 23.0114264 & 0.612\\
138.7487849+29.4746605 & 587738946666758726 & 9:14:59.708 & 29:28:28.778 & 19.9382 & 24.0572414 & 0.498\\
138.7539734+29.7394633 & 588017978339164239 & 9:15:00.954 & 29:44:22.068 & 20.0661 & 22.5720023 & 0.563\\
138.7563217+31.0219574 & 587738948277699044 & 9:15:01.517 & 31:01:19.047 & 20.3730 & 21.775373 & 0.385\\
138.7582448+30.9727119 & 587738948277698992 & 9:15:01.979 & 30:58:21.763 & 20.1276 & 23.2405294 & 0.295\\
138.7623082+29.7296108 & 588017978339164237 & 9:15:02.954 & 29:43:46.599 & 19.5096 & 22.6763703 & 0.389\\
138.7648117+31.0037925 & 587738948277698637 & 9:15:03.555 & 31:00:13.653 & 20.5552 & 22.4070289 & 0.622\\
\enddata
\tablenotetext{a}{Table 2 is published in its entirety in the electronic version of the
{\it Astronomical Journal.} The first ten lines are shown here for guidance regarding its
form and content.}
\tablenotetext{b} {The $g-r$ color is the SDSS fiber color.}
\label{tbl:candidates}
\end{deluxetable}

\clearpage
\begin{deluxetable}{ccccc}
\tablecolumns{5}
\tablewidth{0pc}
\tabletypesize{\footnotesize}
\tablenum{3}
\tablecaption{Luminosity Functions}
\tablehead{
\colhead{} & \colhead{All} & \colhead{HSB} & \colhead{LSB} & \colhead{$SB_{50,R}>22.5$}
}
\startdata
$N$&532&266&266&135\\
$\alpha$ &  $-1.31\pm 0.04$ & $-0.69\pm 0.07$ & $-1.57\pm 0.09$ & $-1.52\pm 0.16$ \\
$M^*_R - 5{\rm log}h$ &  $ -21.32\pm   0.30$ & $ -20.42\pm   0.15$ & $ -18.72\pm   0.34$ & $ -17.32\pm   0.3
4$  \\
$\phi^*/10^{-3}$ mag$^{-1}$ $h^3$ Mpc$^{-3}$ & $ 4.22\pm 0.96$ & $11.7\pm 1.6$ & $ 4.7\pm 2.
0$ & $ 8.5\pm 4.4$   \\
$\chi_\nu^2$  & 1.28 &  0.71 &  1.34 &  0.85 \\
$\nu$  & 16 &  14 &  11 &  9 \\
${\cal L}/10^8 hL_\odot$~Mpc$^{-3}$ & $  0.96\pm  0.35$ & $  0.80\pm  0.25$ & $  0.15\pm  0.
08 $  & $  0.07\pm  0.04 $  \\
\enddata
\label{tbl:lf}
\end{deluxetable}

\clearpage

\clearpage

\clearpage

\clearpage

\clearpage

\clearpage

\clearpage

\clearpage

\clearpage

\clearpage

\clearpage

\clearpage

\clearpage


\clearpage
\begin{thebibliography}{99}

\bibitem[Andreon 
\& Hurn(2010)]{Andreon10} Andreon, S., \& Hurn, M.~A.\ 2010, \mnras, 404, 1922 



\bibitem[Balogh et al.(1999)]{Balogh99} Balogh, M.~L., Morris, 
S.~L., Yee, H.~K.~C., Carlberg, R.~G., 
\& Ellingson, E.\ 1999, \apj, 527, 54 




\bibitem[Benson et al.(2002)]{Benson02} Benson, A.~J., Lacey, 
C.~G., Baugh, C.~M., Cole, S., \& Frenk, C.~S.\ 2002, \mnras, 333, 156 




\bibitem[Benson et al.(2003)]{Benson03} Benson, A.~J., Bower, 
R.~G., Frenk, C.~S., Lacey, C.~G., Baugh, C.~M., 
\& Cole, S.\ 2003, \apj, 599, 38 

\bibitem[Blanton et al.(2005a)]{blanton05a} Blanton, M.~R., et al.\ 
2005a, \aj, 129, 2562 



\bibitem[Blanton et al.(2005b)]{Blanton05b} Blanton, M.~R., Lupton, 
R.~H., Schlegel, D.~J., Strauss, M.~A., Brinkmann, J., Fukugita, M., 
\& Loveday, J.\ 2005b, \apj, 631, 208 

\bibitem[Christlein 
\& Zabludoff(2003)]{Christlein03} Christlein, D., \& Zabludoff, A.~I.\ 2003, \apj, 591, 764 

\bibitem[Cross 
\& Driver(2002)]{2002MNRAS.329..579C} Cross, N., \& Driver, S.~P.\ 2002, \mnras, 329, 579 


\bibitem[de Jong 
\& Lacey(2000)]{2000ApJ...545..781D} de Jong, R.~S., \& Lacey, C.\ 2000, \apj, 545, 781 



\bibitem[Disney 
\& Phillipps(1983)]{Disney83} Disney, M., \& Phillipps, S.\ 1983, \mnras, 205, 1253 

\bibitem[Driver et al.(2005)]{2005MNRAS.360...81D} Driver, S.~P., Liske, 
J., Cross, N.~J.~G., De Propris, R., \& Allen, P.~D.\ 2005, \mnras, 360, 81 




\bibitem[Efstathiou et al.(1988)]{EEP} Efstathiou, G., 
Ellis, R.~S., \& Peterson, B.~A.\ 1988, \mnras, 232, 431 

\bibitem[Fabricant et al.(1998)]{Fabricant98} Fabricant, D.~G., 
Hertz, E.~N., Szentgyorgyi, A.~H., Fata, R.~G., Roll, J.~B., 
\& Zajac, J.~M.\ 1998, \procspie, 3355, 285 



\bibitem[Fabricant et al.(2005)]{Fabricant05} Fabricant, D., et 
al.\ 2005, \pasp, 117, 1411 

\bibitem[Fabricant et al.(2008)]{Fabricant08} Fabricant, D.~G., 
Kurtz, M.~J., Geller, M.~J., Caldwell, N., Woods, D., 
\& Dell'Antonio, I.\ 2008, \pasp, 120, 1222 

\bibitem[Geller et al.(2005)]{Geller05} Geller, M.~J., 
Dell'Antonio, I.~P., Kurtz, M.~J., Ramella, M., Fabricant, D.~G., Caldwell, 
N., Tyson, J.~A., \& Wittman, D.\ 2005, \apjl, 635, L125 

\bibitem[Geller et al.(2010)]{Geller10} Geller, M.~J., Kurtz, 
M.~J., Dell'Antonio, I.~P., Ramella, M., 
\& Fabricant, D.~G.\ 2010, \apj, 709, 832 






\bibitem[Grebel et al.(2003)]{Grebel03} Grebel, E.~K., 
Gallagher, J.~S., III, \& Harbeck, D.\ 2003, \aj, 125, 1926 




\bibitem[Impey et al.(1996)]{1996ApJS..105..209I} Impey, C.~D., Sprayberry, 
D., Irwin, M.~J., \& Bothun, G.~D.\ 1996, \apjs, 105, 209 

\bibitem[Kauffmann et al.(2003)]{Kauffamnn03} Kauffmann, G., et 
al.\ 2003, \mnras, 341, 33 


\bibitem[Kurtz et al.(2007)]{2007AJ....134.1360K} Kurtz, M.~J., Geller, 
M.~J., Fabricant, D.~G., Wyatt, W.~F., 
\& Dell'Antonio, I.~P.\ 2007, \aj, 134, 1360 




\bibitem[Kurtz 
\& Mink(1998)]{KM98} Kurtz, M.~J., \& Mink, D.~J.\ 1998, \pasp, 110, 934 



\bibitem[Liu et al.(2008)]{Liu08} Liu, C.~T., Capak, P., 
Mobasher, B., Paglione, T.~A.~D., Rich, R.~M., Scoville, N.~Z., Tribiano, 
S.~M., \& Tyson, N.~D.\ 2008, \apj, 672, 198 



\bibitem[Lynden-Bell(1971)]{Lynden-Bell71} Lynden-Bell, D.\ 1971, 
\mnras, 155, 95 

\bibitem[Madgwick et al.(2002)]{2002MNRAS.333..133M} Madgwick, D.~S., 
Lahav, O., Baldry, I.~K., et al.\ 2002, \mnras, 333, 133 



\bibitem[Mahdavi et al.(2005)]{Mahdavi05} Mahdavi, A., Trentham, 
N., \& Tully, R.~B.\ 2005, \aj, 130, 1502 



\bibitem[Mateo(1998)]{Mateo98} Mateo, M.~L.\ 1998, \araa, 36, 435 

\bibitem[McGaugh(1996)]{McGaugh96} McGaugh, S.~S.\ 1996, \mnras, 
280, 337 



\bibitem[Mink et al.(2007)]{Mink07} Mink, D.~J., Wyatt, W.~F., 
Caldwell, N., Conroy, M.~A., Furesz, G., 
\& Tokarz, S.~P.\ 2007, Astronomical Data Analysis Software and Systems XVI, 376, 249 

\bibitem[Mobasher et al.(2007)]{Mobasher07} Mobasher, B., et al.\ 
2007, \apjs, 172, 117 


\bibitem[Muller et al.(1998)]{muller98} Muller, G.~P., Reed, R., 
Armandroff, T., Boroson, T.~A., 
\& Jacoby, G.~H.\ 1998, \procspie, 3355, 577 



\bibitem[O'Neil(2002)]{Oneil02} O'Neil, K.\ 2002, Extragalactic 
Gas at Low Redshift, 254, 202 

\bibitem[Roll et al.(1998)]{Roll98} Roll, J.~B., Fabricant, 
D.~G., \& McLeod, B.~A.\ 1998, \procspie, 3355, 324 

\bibitem[Rines 
\& Geller(2008)]{Rines08} Rines, K., \& Geller, M.~J.\ 2008, \aj, 135, 1837 


\bibitem[Sandage et al.(1979)]{1979ApJ...232..352S} Sandage, A., Tammann, 
G.~A., \& Yahil, A.\ 1979, \apj, 232, 352 


\bibitem[Schechter(1976)]{Schechter76} Schechter, P.\ 1976, \apj, 
203, 297 

\bibitem[Sprayberry et al.(1997)]{Sprayberry97} Sprayberry, D., 
Impey, C.~D., Irwin, M.~J., \& Bothun, G.~D.\ 1997, \apj, 482, 104 

\bibitem[Takeuchi et al.(2000)]{2000ApJS..129....1T} Takeuchi, T.~T., 
Yoshikawa, K., \& Ishii, T.~T.\ 2000, \apjs, 129, 1 

\bibitem[Westra et al.(2010)]{Westra10} Westra, E., Geller, 
M.~J., Kurtz, M.~J., Fabricant, D.~G., 
\& Dell'Antonio, I.\ 2010, \pasp, 122, 1258 



\bibitem[White 
\& Rees(1978)]{White78} White, S.~D.~M., \& Rees, M.~J.\ 1978, \mnras, 183, 341 

\bibitem[Wittman et al.(2006)]{Wittman06} Wittman, D., 
Dell'Antonio, I.~P., Hughes, J.~P., Margoniner, V.~E., Tyson, J.~A., Cohen, 
J.~G., \& Norman, D.\ 2006, \apj, 643, 128 

\bibitem[Woods et al.(2010)]{Woods10} Woods, 
D., Geller, M.~J., Kurtz, M.~J., Westra, E., Fabricant, D.~G., 
\& Dell'Antonio, I.\ 2010, \aj, 139, 1857 
\end{thebibliography}
\end{document}